\documentclass{article}
\usepackage{arxiv}
\usepackage[utf8]{inputenc} 
\usepackage[T1]{fontenc}    
\usepackage{url}            
\usepackage{booktabs}       
\usepackage{amsfonts}       
\usepackage{nicefrac}       
\usepackage{microtype}      
\usepackage{graphicx}
\usepackage{epstopdf, epsfig}
\usepackage{amsmath,amssymb}
\usepackage{lineno}
\usepackage{comment}

\usepackage{url}
\usepackage{listings}
\usepackage{lscape} 
\usepackage{longtable}
\usepackage{subfig}
\usepackage[subfigure]{tocloft}
\usepackage[acronym]{glossaries}
\usepackage{authblk}
\usepackage{hyperref}
\usepackage{blindtext}

\usepackage{color}
\usepackage{xcolor}

\newacronym{noaa}{NOAA}{National Oceanic and Atmospheric Administration}
\newacronym{ibe}{IBE}{Inverse Barometer Effect}
\newacronym{hpa}{hPa}{hectopascal}
\newacronym{ecmwf}{ECMWF}{European Centre for Medium Range Weather Forecasts}
\newacronym{ifs}{IFS}{Integrated Forecasting System}
\newacronym{ssh}{SSH}{Sea Surface Height}
\newacronym{bodc}{BODC}{British Oceanography Data Centre}
\newacronym{bfg}{BFG}{Bundesanstalt fũr Gewãsserkunde}
\newacronym{netcdf}{NetCDF}{Network Common Data Form}
\newacronym{nn}{NN}{Neural Network}
\newacronym{ml}{ML}{Machine Learning}
\newacronym{eps}{EPS}{Ensemble Prediction System}
\newacronym{osc}{OSC}{Oscarsborg}
\newacronym{bgo}{BGO}{Bergen}
\newacronym{anx}{ANX}{Andenes}
\newacronym{roms}{ROMS}{Regional Ocean Modeling System}
\newacronym{mse}{MSE}{Mean Square Error}
\newacronym{rmse}{RMSE}{Root Mean Square Error}
\newacronym{adam}{Adam}{Adaptive Moment Estimation}
\newacronym{relu}{ReLU}{Rectified Linear Unit}
\newacronym{nora3}{NORA3}{NORwegian hindcast Archive}
\newacronym{metno}{MET Norway}{Norwegian Meteorological Institute}
\newacronym{ipcc}{IPCC}{Intergovernmental Panel on Climate Change}
\newacronym{dnn}{DNN}{Deep Neural Network}
\newacronym{api}{API}{Application Programming Interface}
\newacronym{meps}{MEPS}{MetCoOp-Ensemble Prediction System}
\newacronym{metcoop}{MetCoOp}{Meteorological Co-operation on Operational NWP}
\newacronym{nwp}{NWP}{Numerical Weather Prediction}
\newacronym{nrt}{NRT}{Near Real-Time}
\newacronym{mae}{MAE}{Mean Absolute Error}

\title{Bias Correction of Operational Storm Surge Forecasts using Neural Networks}

\begin{document}

\author{Paulina Tedesco $^{1, 2, *}$, Jean Rabault $^{2}$, Martin Lilleeng S\ae{}tra $^{3, 4}$, Nils Melsom Kristensen $^{3}$, Ole Johan Aarnes $^{5}$, \O{}yvind Breivik $^{3, 6}$, Cecilie Mauritzen $^{3}$, \O{}yvind S\ae{}tra $^{3}$}

\affil{$^{1}$Department of Physics, University of Oslo, P.O box 1048, Blindern, 0316 Oslo, Norway}
\affil{$^{2}$ Information Technology Department, Norwegian Meteorological Institute}
\affil{$^{3}$ Research and Development Department, Norwegian Meteorological Institute}
\affil{$^{4}$ Department of Computer Science, Oslo Metropolitan University}
\affil{$^{5}$ Equinor ASA, Bergen}
\affil{$^{6}$ Geophysical Institute, University of Bergen}
\affil{$^{*}$ Corresponding author:  paulina.tedesco@gmail.com}

\maketitle


\begin{abstract}

Storm surges can give rise to extreme floods in coastal areas. The \acrfull{metno} produces 120-hour regional operational storm surge forecasts along the coast of Norway based on the \acrfull{roms}, using a model setup called Nordic4-SS. Despite advances in the development of models and computational capabilities, forecast errors remain large enough to impact response measures and issued alerts, in particular, during the strongest storm events. Reducing these errors will positively impact the efficiency of the warning systems while minimizing efforts and resources spent on mitigation. Here, we investigate how forecasts can be improved with residual learning, i.e., training data-driven models to predict the residuals in forecasts from Nordic4-SS. A simple error mapping technique and a more sophisticated \acrfull{nn} method are tested. The simple error mapping technique provides a reduction in the \acrfull{rmse} of less than 4\%. Using the \acrshort{nn} residual correction method, the \acrshort{rmse} in the Oslo Fjord is reduced by 36\% for lead times of one hour, 9\% for 24 hours, and 5\% for 60 hours. Therefore, the residual \acrshort{nn} method is a promising direction for correcting storm surge forecasts, especially on short timescales. Moreover, it is well adapted to being deployed operationally, as i) the correction is applied on top of the existing model and requires no changes to it, ii) all predictors used for \acrshort{nn} inference are already available operationally, iii) prediction by the \acrshort{nn}s is very fast, typically a few seconds per station, and iv) the \acrshort{nn} correction can be provided to a human expert who may inspect it, compare it with the model output, and see how much correction is brought by the \acrshort{nn}, allowing to capitalize on human expertise as a quality validation of the \acrshort{nn} output. While no changes to the hydrodynamic model are necessary to calibrate the neural networks, they are specific to a given model and must be recalibrated when the numerical models are updated.

\end{abstract}

\section{Introduction}
\label{sec:intro}

The \acrfull{ssh} oscillates around the mean sea level following a tidal component and a non-tidal component, i.e. the meteorological component, also called storm surge \citep{pugh_1987}. Tides are regular and periodic sea-level variations with high predictability for the gravitationally driven portion of the water level variability. For the most part, they are directly related to periodical geophysical forcings, such as a combination of the gravitational forces exerted primarily by the Sun and the Moon, and the effect of Earth's rotation and the associated Coriolis force. Tides are commonly the largest source of short-term \acrshort{ssh} fluctuations with a fortnightly neap-spring cycle due to the relative positions of the Sun and the Moon. Their amplitude also varies with the time of the year -- with the strongest tides appearing around the equinoxes due to the alignment of the Sun and the Moon. On the other hand, the principal irregular factors that affect the \acrshort{ssh} are atmospheric pressure and winds acting on the oceans, as well as the associated large-scale waves and currents these trigger. Thus, the relative importance of these two components, tidal and meteorological, depends on the time of the year, weather conditions, and the local bathymetry. For instance, the meteorological component at high latitudes around Norway is greatest during the stormy winter months, particularly over shallow seas.

\begin{figure}
    \centering
    \includegraphics[scale=0.4]{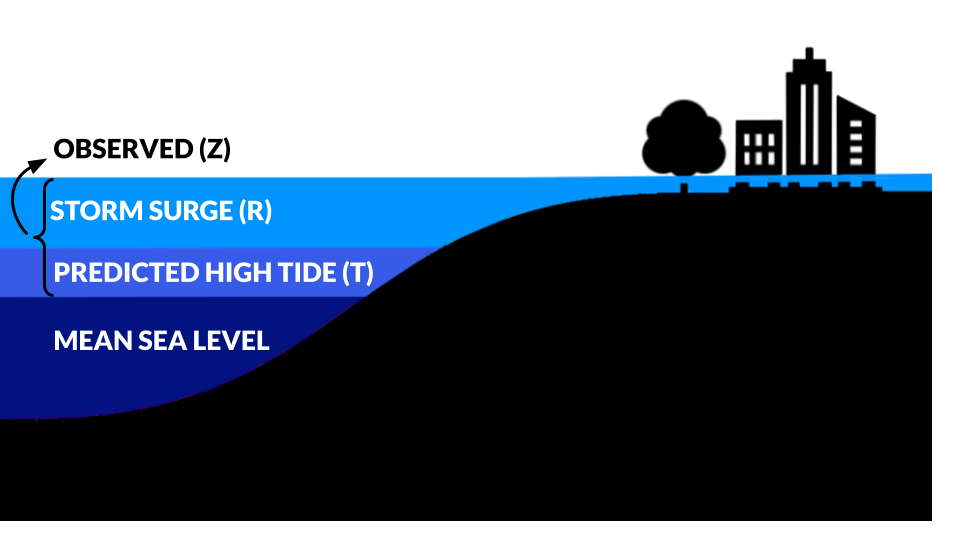}
    \caption{Illustration that shows water level differences for storm surge and a normal (predicted) high tide as compared to mean sea level. Z it the observed total water level, T is the predicted tide with harmonic analysis, and R is the storm surge predicted by the physical model.}
    \label{fig:surge_def}
\end{figure}

Storm surges are formally defined as the height of water above the normal predicted tide, i.e., the meteorological component, see Fig. \ref{fig:surge_def}. However, the term storm surge is usually reserved for events that give rise to unusually high \acrshort{ssh}, rather than minor deviations from the predicted tide. Storm surges can lead to hazardous situations if combined with high tides. For example, when a cyclone makes landfall during a high tide, even of moderate amplitude, it can create an exceptionally high water level rise \citep{pugh_1987}. Similarly, severe storms acting on shallow waters that coincide with spring tides can lead to critical coastal floodings \citep[e.g.,][]{wood_2001}, and potentially damage infrastructure. If the surrounding land is low-lying and densely populated, surges can pose a great threat to life and property \citep[e.g.,][]{gill_1982, pugh_1987, ipccar6}. Furthermore, the soil may become infertile for several years after an inundation because of saline deposits \citep{pugh_1987}. Overall, flood events in populated coastal areas can affect residents’ health, food security, and access to clean water.

From a climate change perspective, impacts and risks are becoming more complex and difficult to manage because of the simultaneous occurrence of multiple hazards \citep{ipccar6}. A combination of, for instance, increased sea level, storm surge, and heavy rain can lead to an extreme event, even if each of these events is not extreme. Above $1.5 ^\circ\mathrm{C}$ global warming, there is high confidence that a combination of these events will increase the compound flood risks \citep{ipccar6}. Given that extreme surge events are projected to increase in the 21st century, the prediction of extreme surge events is a primary concern for designing warning systems and coastal defenses \citep{ipccar6}. For all these reasons, improving the numerical models used to predict storm surges is presently a research area of large interest.

Coastal floods due to storm surges are also a threat to countries around the North Sea, including Norway. Lives were lost in the extreme event of 1960 \citep{kristensen_et_al_2022}, and, although more recent events have not caused deaths, they have damaged properties and caused significant economic losses. For example, the storm surge event that affected Norway in October 1987 and, more recently, the events “Dagmar” (December 2011) and “Elsa” (February 2020). To mitigate potential damages, it is crucial to have a warning system for extreme water levels in place. The \acrfull{metno} has developed a system that predicts and issues warnings in case of extreme levels at 23 permanent Norwegian stations \citep{kristensen_et_al_2022}. Observations at these locations are transferred in \acrfull{nrt} from the Norwegian Mapping Authorities and used for post-processing the forecasts. Then, the corrected values are transmitted from the Research and Development Department to the Forecasting Center at \acrshort{metno}, and published on the official website for water level forecasts operated by the Norwegian Mapping Authorities. Furthermore, a decision support system dashboard for automatically detecting and visualizing the exceedance of certain water level thresholds is internally available to the forecasters at \acrshort{metno}, who will communicate any warnings to key users and the general public \citep{kristensen_et_al_2022}. The core of \acrshort{metno}'s complex warning system is the numerical model, the \acrfull{roms}, which predicts the meteorological component. The astronomical tides computed with harmonic analysis are then added to the storm surge signal to obtain the total water level. In addition to the numerical simulations, the flow of \acrshort{nrt} observations and the dissemination of the forecasts to the authorities and the general public are essential components of the warning system \citep{kristensen_et_al_2022}. In the following, we will refer to the current storm surge model that runs at \acrshort{metno}, which already includes a simple weighted difference correction, as "Nordic4-SS". This correction is applied operationally, by removing the sliding average for the error between the observed and predicted storm surge over the last few hours in the past, as explained further down.

Given that the numerical model is the core of the warning system, improving the forecasts will directly impact the ability to reduce the consequences of extreme storm surge events. While analytical models can be used to analyze the main driving mechanisms behind storm surges, numerical models are required to capture the complexity of weather patterns, bathymetry, and the coupling between the ocean and the atmosphere that impact storm surges \citep{lewis_et_al_2013, mcinnes_et_al_2016, pugh_1987}. Statistical methods, e.g., \acrfull{nn}s, are an alternative to the numerical models frequently used to produce flood warnings to alert the population living in risk areas \citep{hoffken_et_al_2020, tiggeloven_et_la_2021}. The two approaches have been compared in the literature \citep{harris_1962, pasquali_2020}. Numerical models, although capable of describing the physical processes involved, take a long time to develop, set up, and run, require high-quality bathymetric data, are computationally demanding compared to the statistical algorithms, and still have biases and errors owing to the complexity of geophysical systems. These physics-based models have, however, high reliability (even if they are not perfect). Data-driven models, on the contrary, use \acrfull{ml} or other statistical methods to determine the relationship between a set of predictors and the target variables. As a consequence, the complexity of these algorithms is limited by the quantity and quality of the historical and operational data available. They are, however, computationally more efficient than pure numerical models. Moreover, as we show in this article, it is possible to combine the best aspects of the two approaches by correcting the numerical model with a data-driven method \citep{pasquali_2020}, obtaining a third approach. 

There has recently been a renewed interest in the field of \acrshort{ml} \citep{dramsch_2020}, while efforts have been made to model storm surges operationally with \acrshort{nn}s as an alternative to hydrodynamic models \citep[e.g., ][]{deoliveira_et_al_2009, kim_et_al_2016, kim_et_al_2019, das_et_al_2011, tadesse_et_al_2020, sztobryn_2003, makarynskyy_2004}. At a global scale, complex \acrshort{ml} models for predicting the meteorological component of \acrshort{ssh} at hourly intervals using \acrshort{nn}s have been developed \citep{bruneau_et_al_2020, tiggeloven_et_la_2021}, achieving results comparable to those from hydrodynamic models. Together, these studies indicate that \acrshort{nn}s are capable of predicting \acrshort{ssh}, although they do not necessarily improve the performance compared to state of the art physics-based numerical models. On the other hand, this third approach has successfully been applied to correct short-term water level predictions with harmonic analysis in the Galveston Bay \citep{cox_et_al_2002}, and to post-process the meteorological component predicted with a numerical model using data-driven methods in the Adriatic Sea \citep{bajo_2010}. 


In the present work, we show that even a state-of-the-art operational model like Nordic4-SS has significant biases that can be corrected using a residual learning approach that combines numerical modeling with \acrshort{nn}s. Furthermore, we do not limit our work to lead times of about a day, as most previous studies do  \citep[e.g., ][]{deoliveira_et_al_2009, kim_et_al_2016, kim_et_al_2019, das_et_al_2011, tadesse_et_al_2020, sztobryn_2003, bruneau_et_al_2020, tiggeloven_et_la_2021}, but we extend the lead time range to sixty hours, showing the impact of our methods when applied to medium-range forecast. In the first part of the paper, we show that the residuals in Nordic4-SS depend on local wind speed and direction, and present clear bias patterns. The bivariate dependency of the average error on wind variables is visualized in polar plots, a technique that has previously been used in this field before \citep{cox_et_al_2002} and in air quality applications \citep[e.g. ][]{brantley_et_al_2013, grange_et_al_2016, carslaw_and_ropkins_2012}. Although removing the average error in the polar plots does not significantly improve the prediction of extreme surge events, they illustrate and quantify statistical relationships between the variables. These dependencies strongly agree with the intuition and experience of the meteorologists on duty. In the second part of the paper, the \acrfull{rmse} in Nordic4-SS is reduced at several stations by applying a residual \acrshort{nn} method, i.e., by subtracting the residuals estimated with \acrshort{nn}s from Nordic4-SS.

The paper proceeds as follows: First, we describe the methods and the data used; then, we present the results for three selected Norwegian harbors, \acrfull{osc}, \acrfull{bgo}, and \acrfull{anx}; and, finally, we provide a summary of our findings, together with a discussion of the results. We also provide further details in the appendices. All material needed to reproduce these results is available in a Github repository (see Appendix \ref{sec:appendix_a}). A table with the coordinates of the stations is provided in Appendix \ref{sec:appendix_b}. Basic storm surge theory is explained in Appendix \ref{sec:appendix_c} and basic Machine Learning concepts used in the study are explained in Appendix \ref{sec:appendix_d}. Lastly, we show two examples of polar plots as a function of wave characteristics in Appendix \ref{sec:appendix_e}. These figures are not included in the main discussion because the wave parameters do not improve the \acrshort{ml} models, but they are shown to illustrate that error patterns are also visible in other quantities.  

\section{Data}
\label{sec:data}

In this section, we describe the datasets used to to validate and correct the numerical storm surge model. We evaluate \acrshort{roms} in hindcast and forecast modes. However, the usefulness of an improved hindcast in an operational context is limited, so we show only the results of correcting the Nordic4-SS forecasts. Moreover, although the methods have been applied to all the permanent Norwegian stations, we show the results for three of them located in different regions: \acrshort{osc}, \acrshort{bgo}, and \acrshort{anx}. 


\subsection{Total water level observations}
\label{sec:observations}

We use \acrshort{ssh} data from 22 stations in mainland Norway operated by the Norwegian Mapping Authority. These data are transferred in real-time to \acrshort{metno} and used to post-process the forecasts, estimate the residuals in the numerical model and validate it, and as input to the \acrshort{nn}s used to reduce the errors in Nordic4-SS. The geographical location of these stations is shown in Fig. \ref{fig:station_location} (see also Table \ref{tab:station_info} in Appendix \ref{sec:appendix_a} for details).  We have grouped the stations based on our physical understanding of the Norwegian climate, how waves propagate in the ocean, and the geography. The three groups consist of the stations located in Skagerrak (blue), the West Coast of Norway (orange), and Northern Norway (green). Furthermore, we exclude the station Ny Ålesund, in Svalbard, as it is subjected to completely different weather dynamics.



\begin{figure*}[ht]
\begin{center}
\includegraphics[width=.7\textwidth]{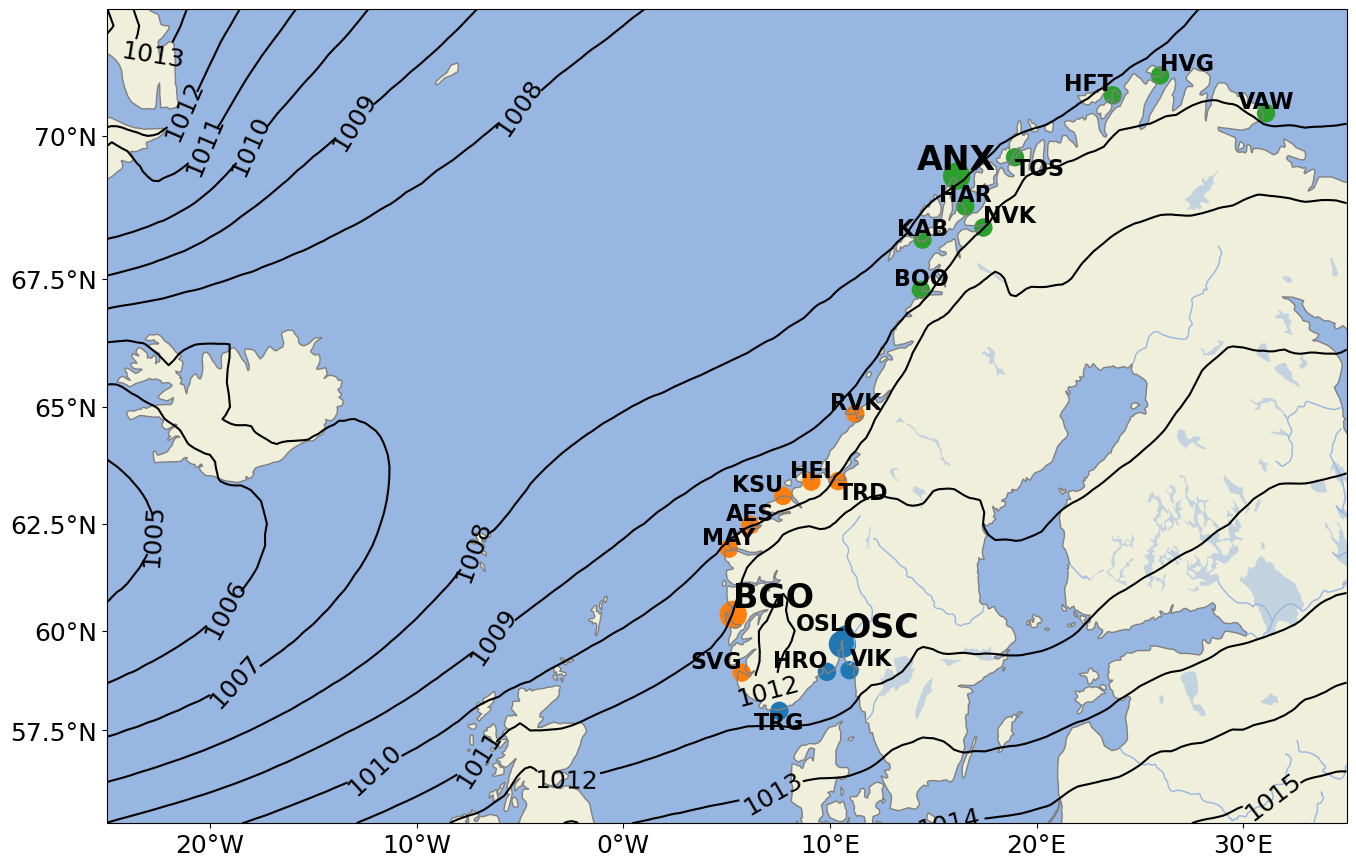}
\caption{\label{fig:station_location}  Locations of the permanent water level stations in mainland Norway. The stations have been divided into three groups according to their geographical locations and the characteristics of tides and weather systems that affect the area. Blue dots represent stations in Skagerrak, orange dots represent stations along the West Coast, and green dots represent stations in Northern Norway. Contours of mean sea level pressure in hPa are computed with data from ERA5 from the period 1959--2021.}
\end{center}
\end{figure*}

\subsection{Tides}
\label{sec:tides}

We used tide data to compute the meteorological component from the observed total water level height, and as predictors in the data-driven models. The tide data used in this work is obtained with harmonic analysis and come from two different sources: a) data retrieved from the Norwegian Mapping Authority's API \citep{api}, and b) estimates made using the pangeo-pytide Python package \citep{pytide}. There are minor differences between the two datasets (in the order of millimeters) that we hypothesize are due to small differences in the number of constituents and the underlying optimization algorithms used by each package. Data from the Norwegian Mapping Authority are preferred, because it is the official source of tide data for Norway, but also because the \acrshort{rmse} for the different models is slightly lower when we train our models on this dataset. Provided that the differences between both datasets are much smaller than the uncertainty in the observations, for convenience, we use a combination of the two datasets depending on what is stored in our systems.


\subsection{Forecasts}
\label{sec:forecast_mode}

Our ultimate goal is to improve the operational storm surge model, Nordic4-SS (01/2018--03/2021). For this, we train \acrshort{nn}s with data that are available at the analysis time, including storm surge and weather forecasts.




\subsubsection{Nordic4-SS surge forecasts}
\label{sec:roms_forecast}
\begin{figure}
    \centering
    \includegraphics[scale=0.15]{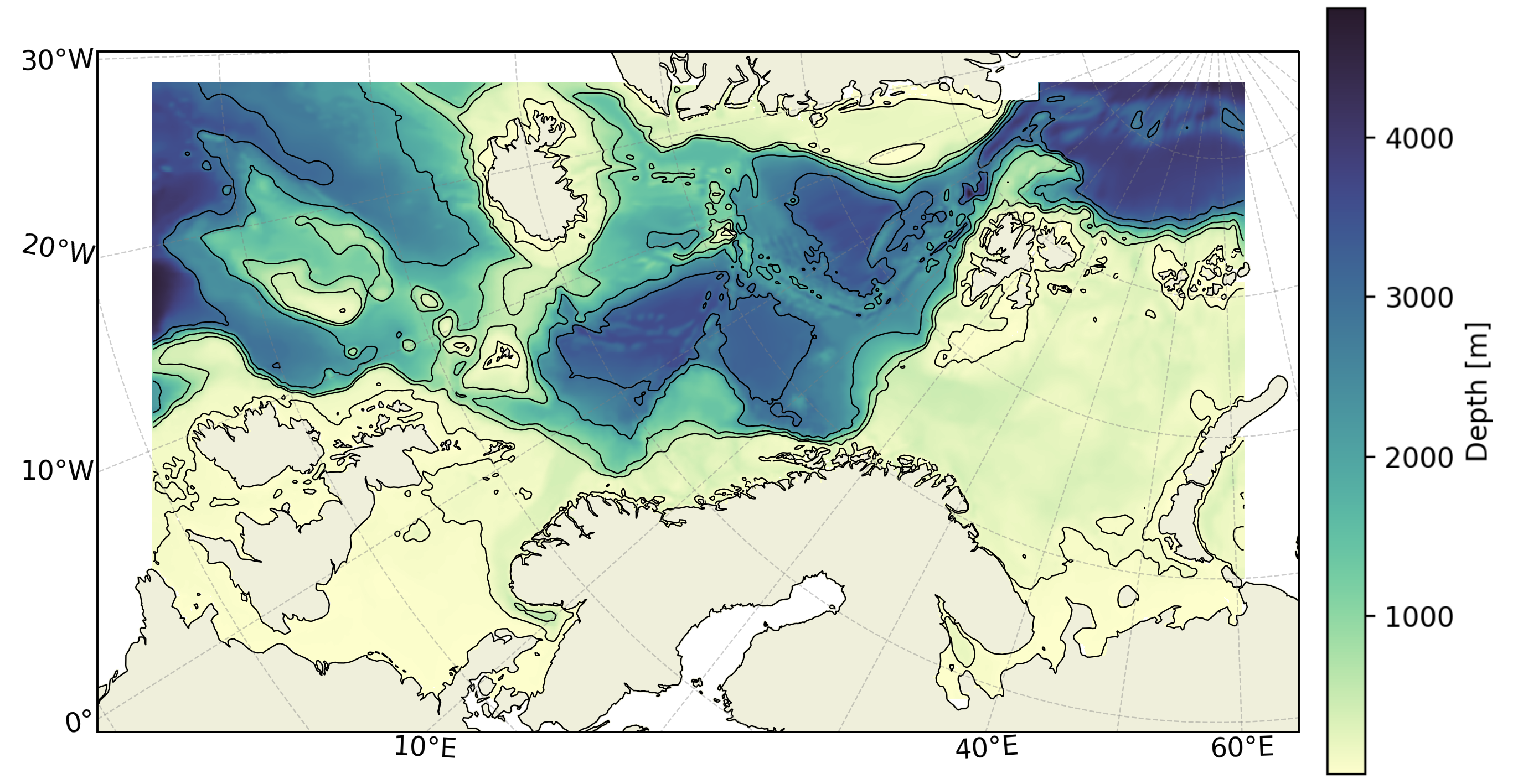}
    \caption{The model domain of Nordic4-SS. The domain has a polar stereographic projection covering the area from Bretagne to Novaya Zemlya, and between Norway and Greenland. Note that the Baltic Sea is not included (masked out) in the model domain. The horizontal model resolution is 4 km. The colors indicate bottom depth in meters, and the contour lines are plotted at 100, 500, 1000, 2000 and 3000 meters. Figure reproduced from "A forecasting and warning system of storm surge events along the Norwegian coast" \citep{kristensen_et_al_2022}.}
    \label{fig:roms_domain}
\end{figure}

The Norwegian storm surge model, Nordic4-SS, is based on \acrshort{roms} \citep{haidvogel_et_al_2008, shchepetkin_and_mcwilliams_2005}, a state-of-the-art model system that has been in operational use, for instance, within the US National Oceanic and Atmospheric Administration for more than a decade, to forecast the water level and currents \citep{haidvogel_et_al_2008}. Nordic4-SS is a free-surface model that uses a terrain-following coordinate system for solving the primitive equations \citep{shchepetkin_and_mcwilliams_2005} with a horizontal resolution of 4 km. The model domain is shown in Fig. \ref{fig:roms_domain}, and covers the North Sea, the Nordic Sea and the Barents Sea. Nordic4-SS consists of the output from ROMS corrected with a weighted difference method (explained in the Section \ref{sec:methods}). These predictions are available for the 23 permanent Norwegian stations through \acrshort{metno}'s Weather \acrfull{api} \citep{api_met}.

\begin{figure}
    \centering
    \includegraphics[scale=0.5]{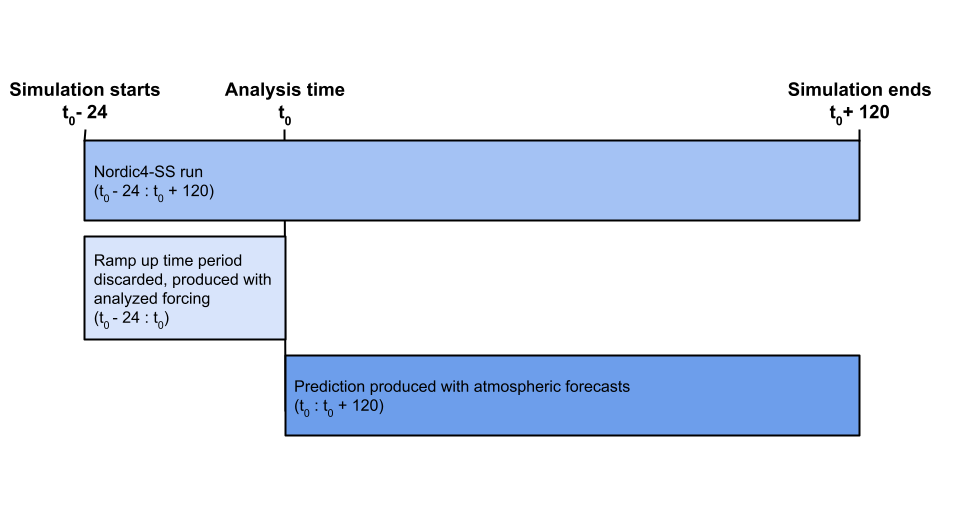}
    \caption{Diagram of Nordic4-SS runs showing the full length of a single run, from 24 hours before the analysis time, $t_0$, to 120 hours after. Each run can be decomposed into a ramp up period before $t_0$, where the model runs with analyzed forcing, and the forecast that starts at $t_0$.}
    \label{fig:nordic4_ss_runs}
\end{figure}

\acrshort{metno} runs \acrshort{roms} in barotropic mode (2D) every 12 hours, and every forecast has a length of 120 hours (five days) \citep{ocean_met, kristensen_et_al_2022} starting from the analysis time. However, for stability reasons, the model run starts 24 hours before the analysis time, meaning that each run is $24 + 120$ hours long (see Fig. \ref{fig:nordic4_ss_runs}). The length of the barotropic time step is set to ten seconds, and the time-stepping scheme followed is the LF-AM3 with Forward Backward (FB) \citep{shchepetkin_and_mcwilliams_2005}. The bathymetry is inherited from a legacy model \citep{engedahl_1995}. Nordic4-SS is forced at the surface boundary with mean sea level pressure and momentum fluxes computed by applying the Charnock relation to 10 meter winds. This forcing data is retrieved from \acrshort{ecmwf} forecasts (note that the first 24 hours of the data are a ramp up time corresponding to 24 hours before the analysis time, as visible in Fig. \ref{fig:nordic4_ss_runs}). The forcing data consists of analysis of pressure and winds from the ECMWF forecasting system. Moreover, a quadratic bottom friction is chosen, with a drag coefficient of $2.5 \times 10^{-3}$. The model runs with open boundary conditions given by the Chapman condition for two-dimensional momentum \citep{chapman_1985} and the Flather condition \citep{flather_1976} for free surface. The values for two-dimensional momentum and sea surface deviation are all set to zero, although the inverted barometer effect is added to the analytic surface deviation at the boundaries.

Nordic4-SS runs in ensemble and deterministic mode. Here, we use only the deterministic model because it is forced by an atmospheric model with higher resolution than the \acrfull{eps} members, which leads to an appreciable reduction in the residuals in Nordic4-SS compared with the other members of the ensemble. 
Furthermore, the length of the dataset used to train the data-driven methods in this study is limited by the archived forecasts of Nordic4-SS, the most recent implementation of the storm surge model, which is available from 2018. 

A fraction of the error in Nordic4-SS is associated with the input to the model. This could be the forcing chosen to run \acrshort{roms}, the description of the bottom topography, or the bottom friction coefficient \citep{haidvogel_et_al_2000, shchepetkin_and_mcwilliams_2005}. Nevertheless, another part of the error is intrinsic to \acrshort{roms} and can be affected by limitations and errors in subscale parameterizations, missing or excluded physics, finite resolution in space and time, discretization and truncation errors \citep{haidvogel_et_al_2000, shchepetkin_and_mcwilliams_2005}, etc. Note that \acrshort{metno} currently runs \acrshort{roms} for storm surge forecasting without tides. However, the storm surge and tidal components are non-linearly dependent and cannot be completely separated for large surges/tides with the magnitude of the impact being a topic of ongoing research \citep[e.g.,][]{williams_et_al_2016, kristensen_et_al_2022}. Some efforts have been made to estimate the intrinsic error in ROMS by running it with analyzed atmospheric forcing. The calculations show that, even though the instrumental error at the water level stations is estimated to be less than 1 cm \citep{kristensen_et_al_2022}, the uncertainties in our ground truth, computed as the difference between total water level and tides, are estimated to be around 3 cm \citep{kristensen_et_al_2022}. This means that additional errors are introduced when, for example, the storm surge signal is derived from the difference between the tides and the total water level. Other potential sources of error in Nordic4-SS are related to local wind effects (as we show in the polar plots in the results section), lack of wave-ocean coupling, or resonances in the basins.

\subsubsection{MEPS meteorological predictions}

The \acrfull{metcoop} is a Nordic cooperation on \acrfull{nwp} between the Finnish Meteorological Institute (FMI), \acrshort{metno}, the Swedish Meteorological and Hydrological Institute (SMHI), and the Estonian Environment Agency (ESTEA). The \acrshort{nn}s developed in this study to improve Nordic4-SS take as input forecasts from \acrfull{meps} \citep{Bengtsson_et_al_2017, frogner_et_al_2019, termonia_et_al_2018}, a forecast ensemble with a convection-permitting atmospheric model covering Scandinavia and the Nordic Seas produced by \acrshort{metcoop}. \acrshort{meps} has a horizontal resolution of 2.5 km and 65 vertical levels. The boundary conditions are taken from ECMWF, and initial perturbations are based on the SLAF method \citep{toth_and_kalnay_1993}. Here, we use \acrshort{meps} forecasts from runs in a 6-hours cycle (00, 06, 12, 18 UTC) with lead times up to 66 hours. Considering that the goal is to design a framework for improving Nordic4-SS forecasts that can be deployed operationally, and that Nordic4-SS runs at 00 and 12, we take the \acrshort{meps} forecasts from the 06 and 18 runs, from lead time 6 to 66. Thus, we design our operational-based setup so as to make sure that the predictions from \acrshort{meps} are available at the analysis time of Nordic4-SS, making it directly transferable to operational applications. This dataset is also used to study the dependence of the residuals in Nordic4-SS in terms of wind speed and direction.

\subsection{Hindcasts}
\label{sec:hindcast_mode}

The storm surge hindcast dataset, NORA3-SS, is used to validate the numerical model because the longer time series it provides give a more detailed insight into the storm surge statistics than the short time series from Nordic4-SS. In this validation process, we also use the atmospheric hindcast data ERA5.

\subsubsection{NORA-SS surge hindcast}
\label{sec:roms_hincast}

\acrshort{nora3} is a high-resolution numerical mesoscale weather simulation made by \acrshort{metno} that covers the North Sea, the Norwegian Sea, and the Barents Sea. It is available from 1974 to 2021 (and will be extended in the future). With a resolution of 3 km, NORA3 downscales the ERA5 reanalysis providing an improved wind field, especially in mountainous areas and along the coastline \citep{haakenstad_et_al_2021, solbrekke_et_al_2021}, and performs much better than ERA5 with regards to the observed maximum wind. The downscaling is based on the HARMONIE-AROME model \citep{muller_et_al_2017_a, muller_et_al_2017_b, Bengtsson_et_al_2017} (Cycle 40h1.2), a nonhydrostatic numerical weather prediction model that explicitly resolves deep convection \citep{haakenstad_et_al_2021, solbrekke_et_al_2021, breivik_et_al_2022}. While the operational storm surge model (Nordic4-SS) is forced with weather forecasts from \acrshort{ecmwf}, the storm surge hindcast (NORA-SS), is forced with reanalysis data from \acrshort{nora3}. Except for the forcing, NORA-SS runs \acrshort{roms} with the same setup as Nordic4-SS.


\subsubsection{ERA5 winds}

In order to study the dependence of the error in NORA-SS on wind conditions, we use gridded reanalysis data from ERA5. Climate reanalyses combine past observations with models to generate time series of climate variables. In this study, the ERA5 reanalysis \citep{hersbach_et_al_2019} was chosen to represent observed historical meteorological and wave conditions, spanning the period 1980--2020.  The regridded data cover the Earth and are available on 37 pressure levels and single levels, on a regular latitude-longitude grid with a $0.25^\circ \times  0.25^\circ$ horizontal resolution and 137 vertical levels. It is also dynamically consistent with the forcing, since NORA3 uses ERA5 as its host model \citep{haakenstad_et_al_2021,haakenstad_et_al_2022}.

Wind data from the gridded datasets are selected from the nearest grid box for each station. Wind speed and direction are calculated from the eastward ($u$) and northward ($v$) components of the wind at ten meters at an hourly frequency. Experiments were also conducted using pressure and wave data, but we focus on the wind dependency because, out of these three, the wind parameters are experimentally found to be the most important \acrshort{nn} predictors for correcting Nordic4-SS.

\section{Methods}
\label{sec:methods}

In this section, we introduce the methodology used to reduce the residuals in \acrshort{metno}'s storm surge model, Nordic4-SS. Although the methodology has been developed for improving the predictions made with Nordic4-SS at stations located along the Norwegian coast, it can easily be generalized to other models and regions as long as in-situ observations and numerical model data are available.  

In the first part of the work, we validate the numerical model using traditional statistical methods (this will be referred to as ``traditional'' in the following). Polar plots are used to display the dependency of the systematic bias, either in Nordic4-SS or NORA-SS, on two variables simultaneously; for instance, local wind speed and wind direction. These plots, and their corresponding tables, can not only be used for validation, but also to correct the average error. In the second part of this work we apply \acrshort{nn}s. Finally, the models are evaluated and compared against each other using the \acrshort{rmse}, and we select the model with the best performance.

\begin{figure}
    \centering
    \includegraphics[scale=0.43]{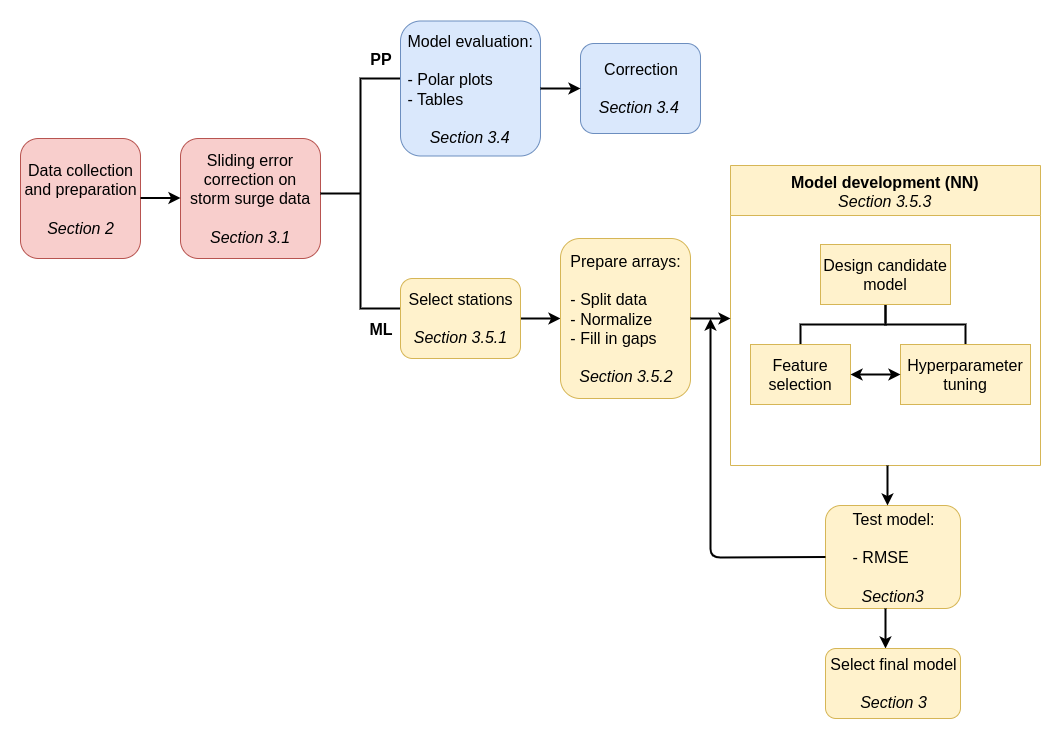}
    \caption{Workflow diagram for validation and correction with the polar plots method (PP) in blue and the correction with the Machine learning method (ML) in yellow. Common for both methods (in red) is the data collection and preparation and the application of the weighted differences correction on the storm surge data. The ML part has several components. We start by selecting a subset of stations from which we extract the predictors and prepare labels and feature arrays. Then, we develop the \acrshort{nn} models, design the architecture, select the features and tune the hyperparameters. Finally, we select the model with the best RMSE computed with test data.}
    \label{fig:workflow}
\end{figure}

The sequence of processes followed to validate and correct Nordic4-SS is represented in the workflow diagram in Fig. \ref{fig:workflow}. The first step consists of collecting and processing all the data needed from the different sources described above. Then, we compute the polar plots and train the \acrshort{nn}s on the residuals. This workflow diagram is explained in detail in the following. The following notation is used in the next sections: $Z$ refers to the observed \acrshort{ssh}, $T$ stands for tide, and $R$ is the meteorological component predicted with \acrshort{roms} (as illustrated in Fig. \ref{fig:surge_def}). 




\subsection{Post-processing the storm surge predictions}

The outputs from Nordic4-SS and NORA-SS are adjusted with the \textit{weighted differences correction method}~\citep{kristensen_et_al_2022}. The method relies on the observations of the previous five days and is applied before computing the residuals needed for validating and correcting the numerical model. At each location, the elements in the offset vector $O$ represent the error from $t-120$ to $t-1$ hours:

\begin{equation}
    e(t) = (Z(t) - T(t)) - R(t),
\end{equation}

\begin{equation}
    O(t) = \begin{bmatrix}
    e(t-120) & e(t-119) & \ldots & e(t-1)
    \end{bmatrix}.
\end{equation}

Then, the bias is computed as the sum of the weighted offsets, where the last observations have larger weights:

\begin{equation}
    W = \begin{bmatrix}
   1 & 2 & \ldots & 120
    \end{bmatrix} / \sum_{i=1}^{120} i,
\end{equation}

\begin{equation}
    \mathrm{bias_{WD}} =  O \times W'.
    \label{eq:bias_met}
\end{equation}

As we show in the following, even after removing the bias computed with this weighted differences correction method, $\mathrm{bias_{WD}}$, there is a systematic error in the \acrshort{roms} output. To further compensate for it, we learn these errors with two different data-driven methods: the traditional method and the \acrshort{nn}s.


\subsection{Residual framework}

The residual errors are defined as the differences between what we expect and what is predicted, in our case, the observed and predicted storm surge. We define the residuals at the location $s$, corresponding to a node on the discrete numerical grid, and time $t$ as:

\begin{equation}
    \epsilon_{s, t} = (Z_{s, t} - T_{s, t}) - (R_{s, t} - \mathrm{bias_{WD}}),
    \label{eq:error_roms}
\end{equation}

where $R_{s, t}$ is the output from \acrshort{roms}, run either in forecast mode (Nordic4-SS) or hindcast mode (NORA-SS), at a given time and location, corrected with the weighted differences method by subtracting $\mathrm{bias_{WD}}$; $Z_{s, t}$ is the total observed height; and $T_{s, t}$ is the tide estimated with harmonic analysis. All values are measured with respect to the official chart datum. Note that the residuals form a time series themselves.

In an ideal model, the residuals should be random fluctuations. Any structure in the residuals suggests that the original model is not perfect and could be improved. When the signal in the residuals is complex, it might be convenient to model the structure directly, i.e., model how the forecasting model will fail, to later remove the predicted residuals from the numerical model and improve its performance. To this end, we use \acrshort{nn}s as a post-processing tool, training \acrshort{nn}s on the signal in the residuals, which is a less complex task than predicting directly the full storm surge dynamics. It is also particularly convenient to model a less complicated signal in the light of short training samples and the limited in situ ground truth data available, which puts a limitation on the complexity of the \acrshort{nn} model that can be used before overfitting.

Autoregressive models are traditionally used to model autocorrelated residuals, where the lagged errors are combined in a classical regression model. However, the fact that \acrshort{nn}s are inherently nonlinear makes them better candidates for modeling complex data patterns than traditional methods. We use the time lag concept from the autoregressive models, but we combine it with a more flexible learning algorithm, \acrshort{nn}s. That is, the input nodes of the \acrshort{nn} consist of time-lagged variables. Although the individual values in the lagged variables are duplicated, the \acrshort{nn} training process can assign unique weights to the vectors with lagged data for each variable while learning how past values influence future values. The forecasting performance of our algorithms is affected by the time lag selection, in addition to the model selection and setup. Therefore, it is essential to select the time lags carefully. We have verified with our data sample that, if the time window is too short, the model will not have enough information to learn the correlations in time. Contrarily, we see that a too large time-lag value results in irrelevant inputs and reduce the performance. In our experiment, the number of inputs, hence, the number of time lags, is limited by the length of the records. Using too many time lags results in a large model size compared with the size of the training dataset available, which leads to overfitting. A sensitivity analysis concluded that 24 hours of lagged values for each variable is optimal given the size of our dataset. By using this range of lagged variables we expect the \acrshort{nn}s to model the highest autocorrelated part of the signals in the residuals and the semi-diurnal periodicity in the tide-surge interaction missing in the system.


\subsection{Training, validation and test}

When using data-driven algorithms to make predictions, it is common to split the data into a training and a test dataset. We fit the parameters to the training dataset. When finding the model with the best performance and adjusting the hyperparameters, it is necessary to set apart a fraction of the training data for validation. This validation dataset is neither part of the low-level training nor the final testing. Once a model is selected, the performance should be assessed on an independent dataset, namely the test dataset. 

\begin{figure}
    \centering
    \includegraphics[scale=0.3]{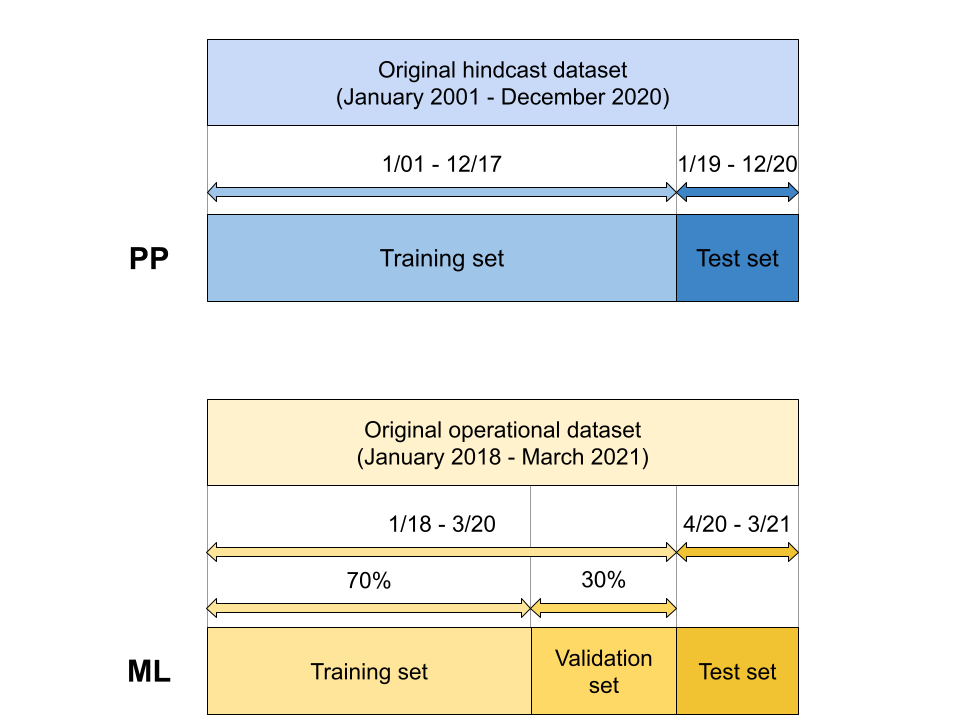}
    \caption{Train-test split of the original dataset for the polar plot (PP) method and the \acrshort{ml} method. In the case of the PP method, the training dataset consists of the hindcast data spanning the period January 2018 -- December 2017, whereas the test dataset consists of two years of data, from January 2019 -- December 2020. The \acrshort{ml} corrects the Nordic4-SS predictions. In this case, the training dataset consists of $70\%$ of the data in the Nordic4-SS data in the period January 2018 -- March 2020, whereas the remaining $30\%$ is held out for validation. The test dataset covers the period April 2020 -- March 2021.}
    \label{fig:train_test_split}
\end{figure}

How we split the datasets in this study depends on whether we are using hindcast or forecast data, and whether we are training traditional methods (polar plots) or \acrshort{nn}s (see Fig. \ref{fig:train_test_split}). The operational test dataset consists of only one year of data, from April 2020 to March 2021. Since we have a small input sample, the year we select for testing unfortunately impacts the results, as there are years with more storm surge events than others. The hindcast dataset, only used for validation of the numerical model with polar plots, is longer. We split it into a training dataset extending from January 2001 to December 2017 and a test dataset from January 2018 to December 2019. We do not need a validation dataset for traditional methods, but for the \acrshort{nn}, we split the training data and leave $70 \%$ for low-level training and $30 \%$ for validation.

\subsection{Statistical bias correction}
\label{sec:statistics_correction}

\begin{figure}
    \centering
    \includegraphics[scale=0.5]{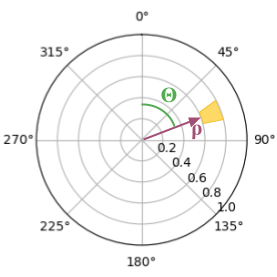}
    \caption{Schematic polar plot representation, where the $\rho$ represents the radial coordinate (e.g., wind speed), $\theta$ represents the angular coordinate (e.g., wind direction), and the yellow square is the binned data (e.g., average residuals).}
    \label{fig:radial_plot_diagram}
\end{figure}
In the first part of this work, we show that the residuals in the storm surge model (Eq.  \ref{eq:error_roms}) depend on local wind conditions and suggest a traditional 2D polar plot method to analyze the joint dependence on wind speed and direction.  In these plots, the wind direction is represented by the angular coordinate, whereas the wind speed is represented by the radial coordinate and increases with the radius (see Fig. \ref{fig:radial_plot_diagram}). Then, we calculate each bin's average error, standard deviation, and number of observations. We divide the data into 2D bins instead of interpolating and plotting a continuous field, and keep only the bins with at least five observations. The optimal bin sizes for the datasets used in this work is 1 m/s $\times$ 10 degrees for the ERA5 wind data. Although polar plots can be used to correct the statistical bias in the storm surge model by computing the average residuals for each bin and then subtracting this bias from Nordic4-SS, they are most useful as a visual representation of the error in polar coordinates in a validation process. 



\subsection{Machine learning}
\label{sec:ml}

The traditional statistical method involving polar plots can potentially be used to remove a part of the systematic error in the storm surge forecast. Nonetheless, it has two major drawbacks: 1) the performance is poor in the case of extreme events, due to the scarcity of rare events in the training dataset, and 2)  because variables are correlated, it can only be used to correct the bias associated with two predictor variables, such as wind speed and direction, for one location at a time. One way of mitigating this problem is to use \acrshort{nn}s. These are more flexible, nonlinear models with the ability to model complex relationships between inputs and outputs. 

In order to reduce the bias in Nordic4-SS, we apply the residual method with \acrshort{nn}s, i.e., we predict the residuals in the numerical model and then subtract them from the storm surge predictions. We apply this method to each station independently. The models have been implemented with the Keras library \citep{keras}, for hourly lead times ranging from one to 60 hours.

\subsubsection{Station selection}
\label{sec:station_selection}
The Norwegian climate and storm surge conditions are affected by the country's geography. A long, intricate coast line, deep and narrow fjords, high mountains, and steep valleys are important factors to consider in prediction systems. Therefore, it is natural to use predictors from different stations depending on where we want to predict the residuals. However, the choice of stations is not trivial. In theory, for each of the 22 stations where we want to improve the forecasts, we could test the performance of the \acrshort{ml} models using all possible combinations of predictor variables and stations, but the number of possible combinations means that a direct testing approach would require enormous efforts and computational resources. A more practical solution involves grouping the 22 stations and selecting a set of predictor stations for each group. The groups have been determined by performing the $k$-means algorithm for $k=3$ on the storm surge data, and coincide with the physical-based groups shown in Fig. \ref{fig:station_location}. The $k$-means method has been tested for different numbers of $k$. For a partition based on $k=4$, the West Coast is divided into two groups of stations, north and south of Bergen, leaving a too small number of stations in one of the groups. In addition, for $k=5$, Northern Norway is divided into two groups. On the contrary, when running the algorithm for $k=2$, the stations are grouped according to their latitude, meaning that most of the stations on the West Coast are grouped with the stations in Skagerrak, which does not agree with the geography and the physics of the basins. In summary, the lowest number of clusters coherent with the geography and the physics is $k=3$, and a higher number of clusters could be considered if we had more in situ observations.

Moreover, when predicting the residuals at a given location, it is useful to provide the \acrshort{nn}s with data from remote locations that contain information about the weather systems at a previous state. Therefore, to select the predictor stations for each of the three groups of stations, we consider how wave and weather systems propagate. We take, for instance, into account that tides are mostly generated in the Atlantic, before a part of the wave propagates along Britain and follows the coast in Northern Europe, reaching Skagerrak, and another part propagates to the Norwegian Sea and continues northwards. Weather systems typically move in a north-easterly direction, but in Southern Norway they can also move along a more zonal path. In addition, we see that stations located in the inner sections of the fjords have particularly complicated dynamics and, as such, the data from these stations are not good candidates as input to the \acrshort{nn}s. Also, due to autocorrelations in the residuals, past observations and forecasts at the station where we want to improve the forecast are key predictors. Furthermore, for robustness, we conduct experiments to test the performance of the \acrshort{ml} algorithms on a subset of possible combinations of stations.



In summary, each group shown in Fig. \ref{fig:station_location} has its corresponding set of predictor stations. All \acrshort{nn} models include data from the station we are predicting at and data from 4 other stations, depending on which group they belong to. We use data from AES, BGO, VIK, and TRG for the stations in Skagerrak (if the station for which we predict the residuals is in this list, we add OSC to complete the set of 5 stations), AES, OSC, MAY, and SVG for the stations along the West Coast (if the station is among the predictors, we add BGO), and BOO, HFT, KAB and KSU for the stations in Northern Norway (if the station is among the predictors, we add ANX). 

\subsubsection{Data preparation}
\label{sec:data_prep}

The \acrshort{nn}s perform better on normalized data. Before training the models, we standardized all the data using the sample mean and the standard deviation computed with the training data. The opposite transformation is then applied to test the models by comparing the \acrshort{rmse}. We tested alternative normalization methods, but they were found to either have no impact on, or degrade, the accuracy of the \acrshort{nn}. The gaps in the data are filled with the training sample's most frequent value, with scikit-learn's SimpleImputer class \citep{pedregosa_et_al_2011}. At OSC, BGO, and ANX, this corresponds to 1.8 \% of the total prediction dataset, after removing missing labels (8.5 \%). If no repeated values are found, the algorithm selects the minimum of the dataset. 

\subsubsection{Model development}
\label{sec:model_dev}

Once we have identified the model architecture we want to use, multiple hyperparameters must be specified before beginning the training process. There is no analytical way to determine the optimal values of these parameters. Instead, we rely on systematic experimentation and testing. However, the optimal hyperparameters will depend on the features selected and the architecture.

\begin{figure}
    \centering
    \includegraphics[scale=0.5]{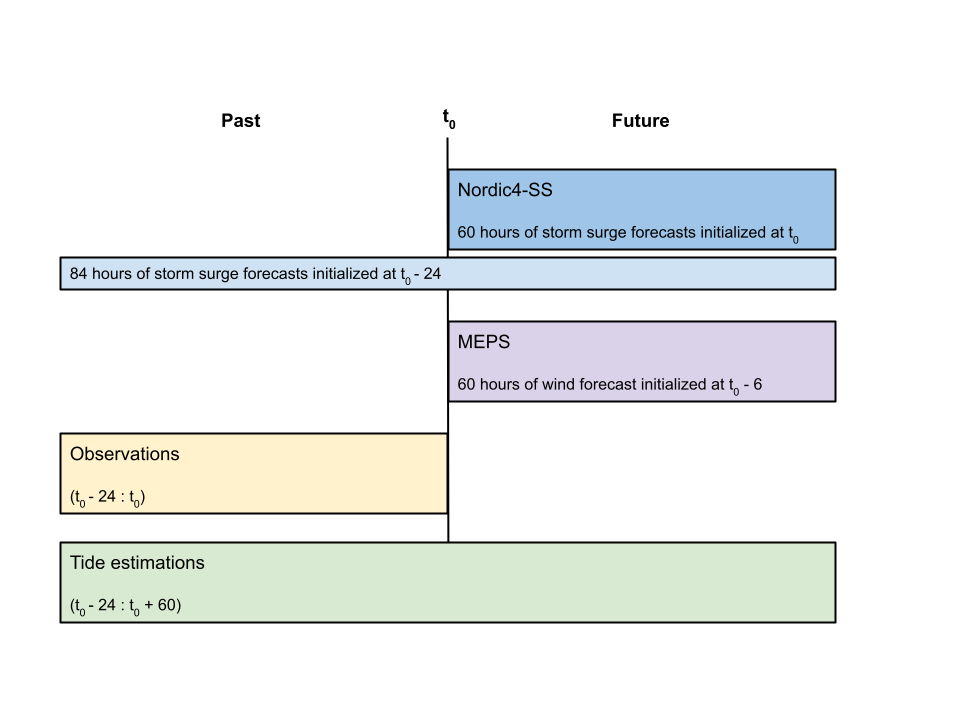}
    \caption{Diagram of predictors used to train the \acrshort{nn}s in forecast mode for lead times up to 60 hours. We use storm surge forecasts from Nordic4-SS generated at the analysis time, $t_0$, but also forecasts from 24 hours before $t_0$. We also use weather forecasts from MEPS generated six hours before the $t_0$ from lead time six to 66 hours, observations from the last 24 hours, and tide estimates for the last 24 hours up to $t_0 + 60$ hours.}
    \label{fig:setup_diagram_forecast}
\end{figure}

The residuals are estimated for one station at a time using input data from several locations, including the one where we want to predict. The number of predictor variables is limited by the length of the Nordic4-SS forecasts. Therefore, we have to carefully select the best candidates, discarding predictors that carry less value. The set of predictor variables consists of observations, Nordic4-SS predictions (initialized at $t_0$ and $t_0 - 24$) corrected with the weighted differences correction method (Eqn.  \ref{eq:bias_met}), tides, and 10 m wind forecast variables from \acrshort{meps} (initialized at $t_0$). The maximum range of hours is from $t-24$ to $t+60$. Note that observations are only provided for the past to make the method forecast compatible. The diagram in Fig. \ref{fig:setup_diagram_forecast} shows the period spanned by each variable relative to the analysis time. 

\begin{figure*}[ht]
  \centering
  {\includegraphics[width=0.4\textwidth]{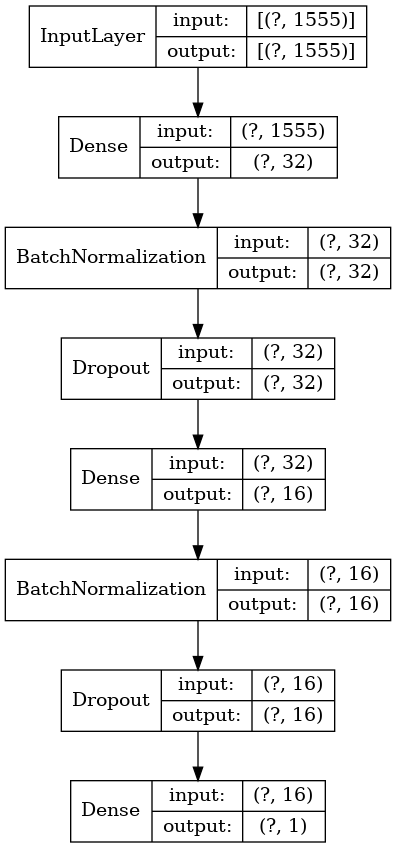}}
  \caption{Illustration of the \acrshort{nn} model graph using the direct strategy. The graph shows the layer order in the model with their corresponding input and output shapes. The number of observations provided to the model is variable, here represented by a question mark. It is assumed that data from 5 locations are used to train the networks; therefore, the number of columns in the input layer is 1555.}
  \label{fig:architecture}
\end{figure*}

 We use a direct multi-step forecasting strategy that involves training a separate model for each forecast time.  The architecture is that of a sequential model, consisting of a dense layer of 32 nodes, followed by a batch normalization layer, a dropout layer with rate 0.3, a dense layer of 16 nodes, another batch normalization layer, and a dropout layer with rate 0.4 (see Fig. \ref{fig:architecture}). Thus, the number of nodes decreases with the layer number, from the first to the last one. The weights in all the dense layers have been initialized with the Glorot uniform initializer \citep{glorot_and_bengio_2010}, and the nodes are activated according to the \acrfull{relu} function.


The architecture of the \acrshort{nn} models used to predict the residuals at each lead time is illustrated in Fig. \ref{fig:architecture}. Each node in the graph represents a specific layer function, while the arrows represent the flow. This way, the \acrshort{nn} graphs show the order of the layers, starting with the input layer on top and ending with the last dense layer (output) at the bottom. The nodes also include the input and output shapes of each layer. The number of predictors is variable, but if we train the models with observations, tide, storm surge, and wind from five different stations, from $t-24$ to $t+60$, after removing predictors with missing data the number of predictors for \acrshort{osc} is 1550 (as shown for the input layer in Fig. \ref{fig:architecture}). Note that the number of observations, or samples, used to train the networks on each operation is variable. Because the \acrshort{nn} operates on a batch of the input, the question marks in the graph are a placeholder for the number of samples. In this work, we have used a batch size of 32. 

The \acrfull{adam} optimizer \citep{kingma_and_ba_2014} is used for performing the training, with a learning rate of 0.001. The advantage of using this method is that it converges rapidly, and no manual tuning of the learning rate is needed. Moreover, we have chosen the \acrfull{mse} as a loss function. In order to evaluate the model, we use the Mean Absolute Error (MAE). This metric is used to judge the model's performance, not in the training process. When the metric has stopped improving for 20 epochs, the learning rate is reduced by a factor of 2. The lower bound of the learning rate is set to 0.0001. The maximum number of iterations is 500, but the training terminates when the loss does not improve over 50 epochs.

\section{Results}
\label{sec:results}

Herein, we analyze the residuals in the numerical model at the Norwegian stations with simple statistical methods, and use \acrshort{nn}s to learn these residuals in order to improve Nordic4-SS. We start by validating the numerical model and searching for a signal in the residuals. We find that the residuals are correlated in time and depend on the wind conditions. As will be shown, when learning these errors, \acrshort{ml} algorithms outperform traditional methods. However, the more complex the algorithm is, the more difficult it is to interpret the results. We therefore believe that the polar plots are a convenient complement to the \acrshort{nn} technique as they expose the physical relationship between the systematic error in the numerical model and the wind variables. Hence, they allow us to better understand the flaws of the storm surge model, and where it fails. Moreover, analyzing the statistics of the error in the numerical model through polar plots is a way to systematize the knowledge obtained through years of experience of the forecasters. Even though the methodology described above is independent of the location, there is an evident spatial variability in our results. We focus on the results obtained for three stations, each of them located in one of the three different Norwegian regions defined above (see Fig. \ref{fig:station_location}): \acrfull{osc} is located in Skagerrak, \acrfull{bgo} in the West Coast, and \acrfull{anx} in Northern Norway.

\subsection{Validation of the numerical storm surge model}

\begin{figure}[ht]
  \centering
  \subfloat[Residuals OSC]{\includegraphics[width=0.28\textwidth]{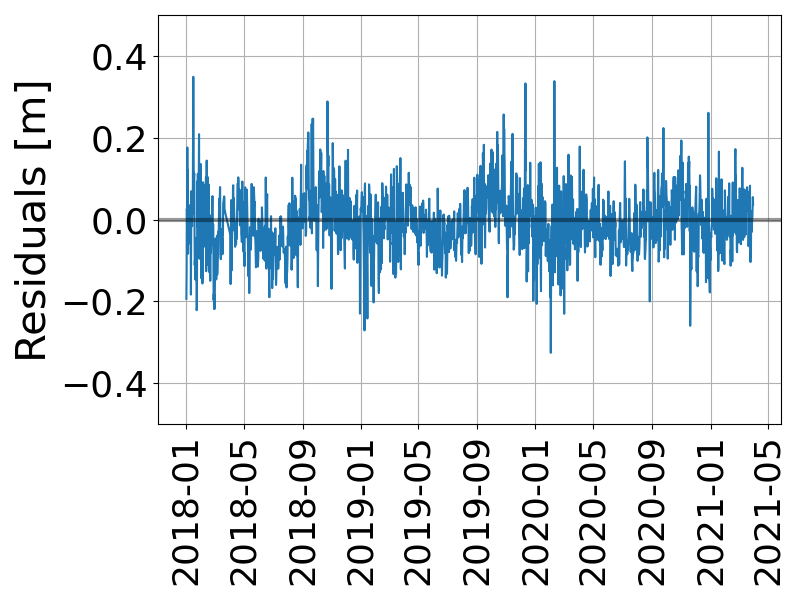}\label{fig:timeseries_res_NO_OSC}}
  \hfill
  \subfloat[Residuals BGO]{\includegraphics[width=0.28\textwidth]{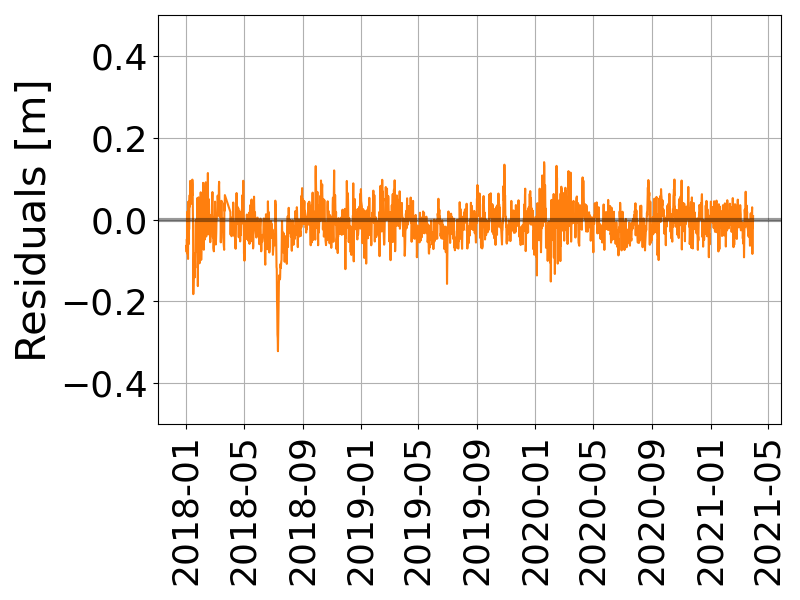}\label{fig:timeseries_res_NO_BGO}}
  \hfill
  \subfloat[Residuals ANX]{\includegraphics[width=0.28\textwidth]{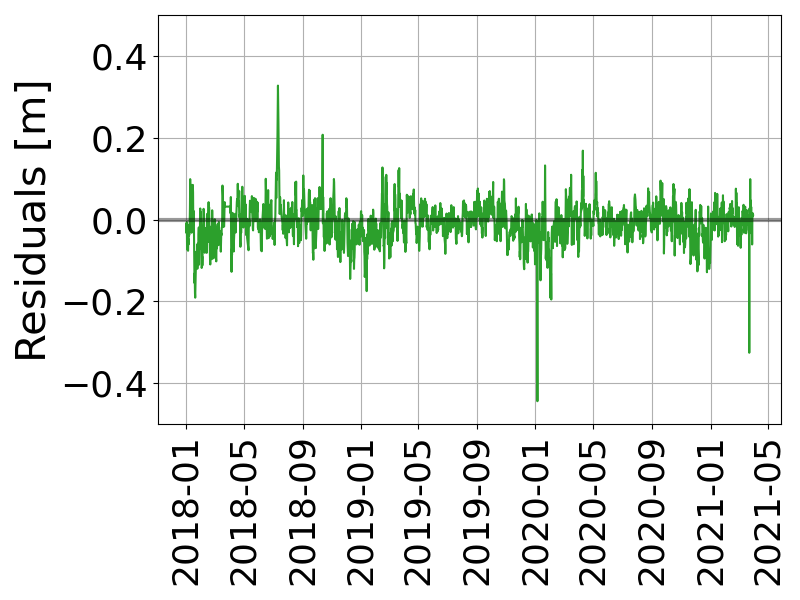}\label{fig:timeseries_res_NO_ANX}}
  
  \subfloat[Histogram residuals OSC]{\includegraphics[width=0.28\textwidth]{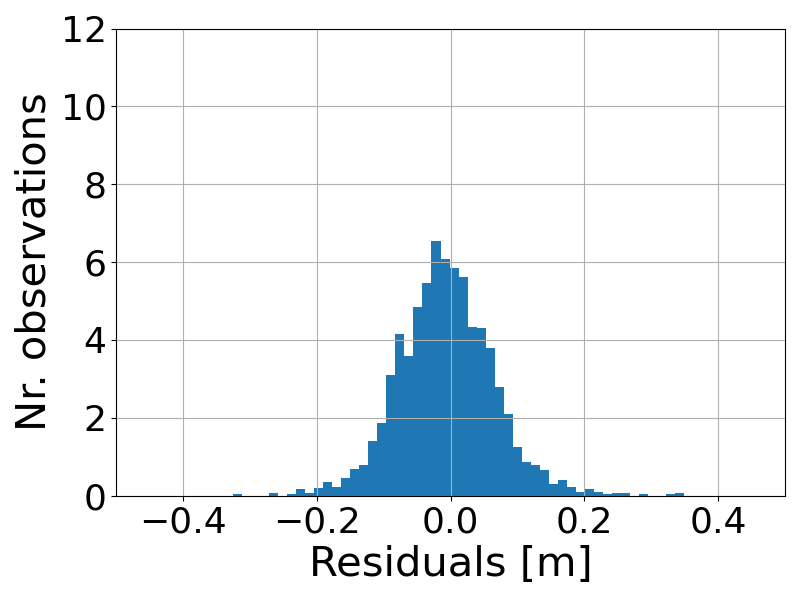}\label{fig:hist_res_NO_OSC}}
  \hfill
  \subfloat[Histogram residuals BGO]{\includegraphics[width=0.28\textwidth]{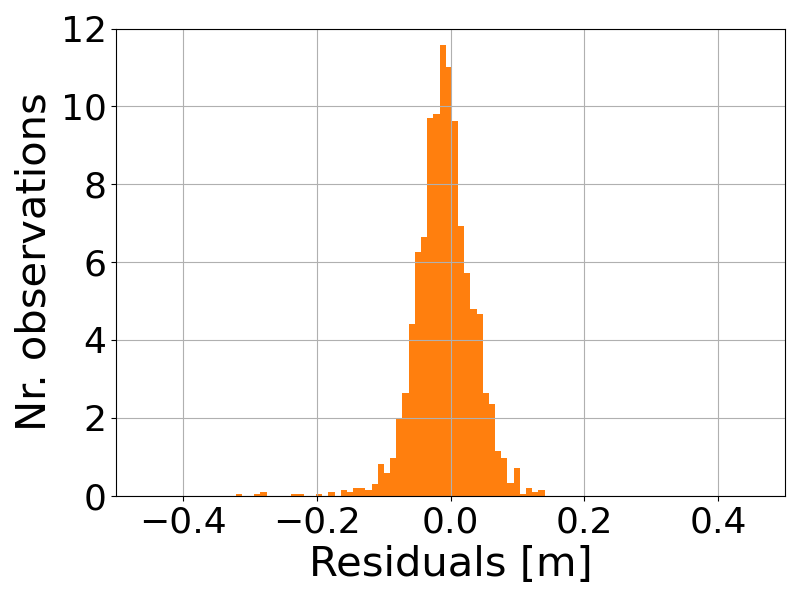}\label{fig:hist_res_NO_BGO}}
  \hfill
  \subfloat[Histogram residuals ANX]{\includegraphics[width=0.28\textwidth]{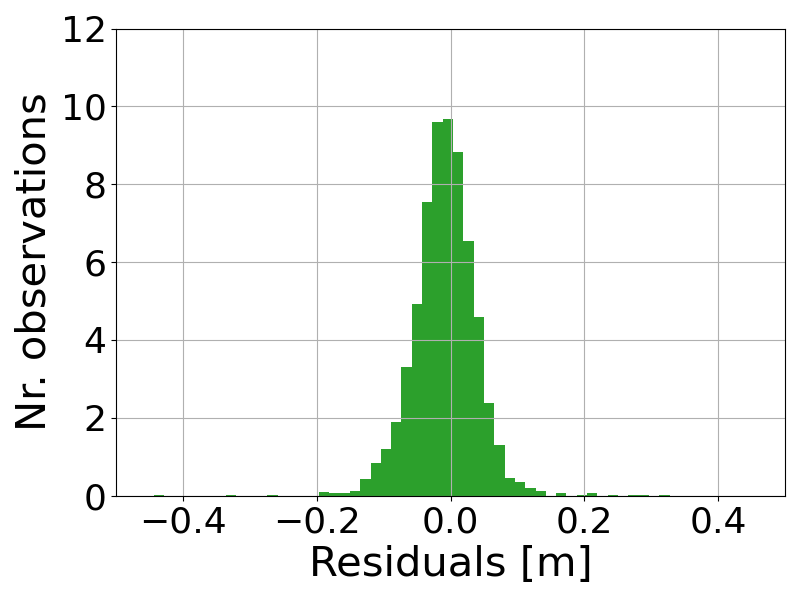}\label{fig:hist_res_NO_ANX}}
  
  \subfloat[Autocorrelation residuals OSC]{\includegraphics[width=0.28\textwidth]{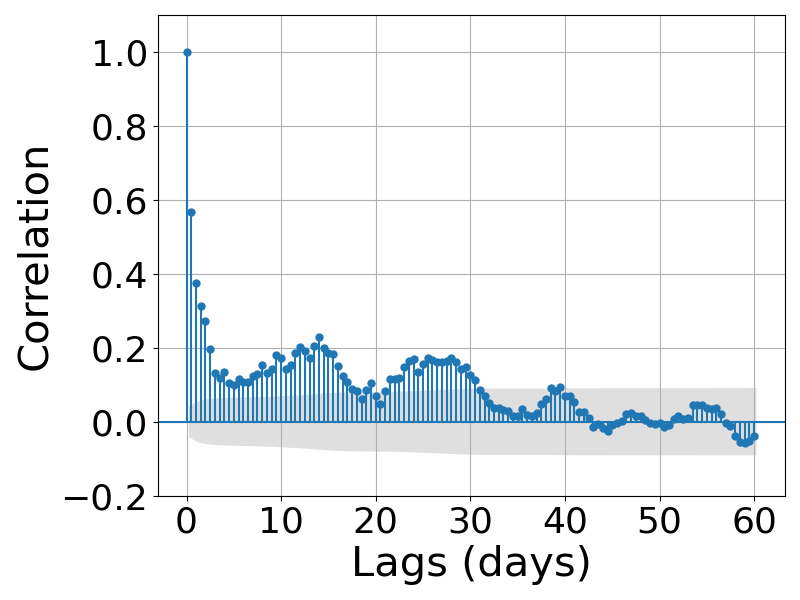}\label{fig:autocorr_res_NO_OSC}}
  \hfill
  \subfloat[Autocorrelation residuals BGO]{\includegraphics[width=0.28\textwidth]{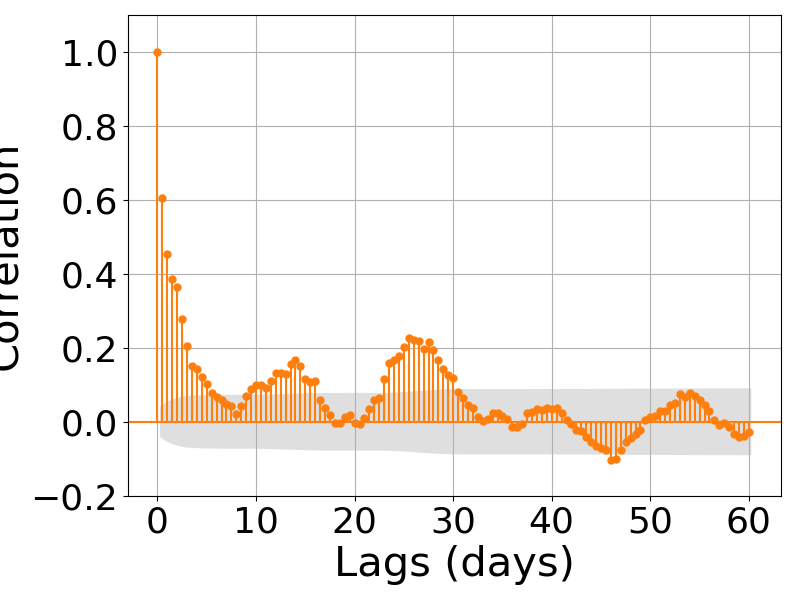}\label{fig:autocorr_res_NO_BGO}}
  \hfill
  \subfloat[Autocorrelation residuals ANX]{\includegraphics[width=0.28\textwidth]{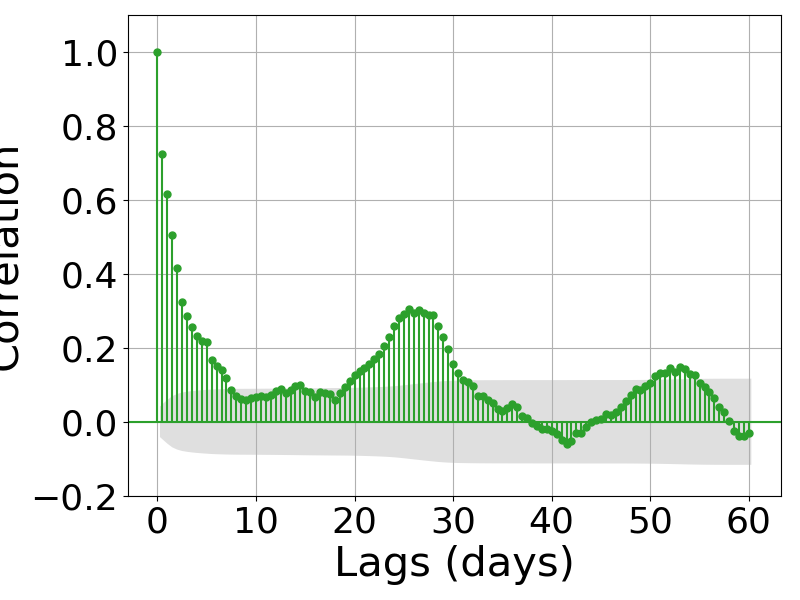}\label{fig:autocorr_res_NO_ANX}}

  \caption{Statistics of the residuals in \acrshort{metno}'s operational storm surge model (Nordic4-SS). The residuals have been computed as the difference between the observed and predicted meteorological component, where the predicted meteorological component is the output from Nordic4-SS corrected with a weighted differences correction method. The panels show the time series of the residuals at a) \acrfull{osc}; b) \acrfull{bgo}; and c) \acrfull{anx}; histograms of the residuals at d) \acrshort{osc}; e) \acrshort{bgo}; and f) \acrshort{anx}; the autocorrelations in the residuals up to 60 days at g) \acrshort{osc}; h) \acrshort{bgo}; and i) \acrshort{anx}. Note that the figures were constructed with 12-hourly data from the period January 2018--March 2021.}
  \label{fig:stat_residuals}
\end{figure}
Fig. \ref{fig:stat_residuals} shows time series, histograms, and autocorrelation plots of the residuals in Nordic4-SS at each of the three chosen locations. Each color represents a station. Although the magnitude of the residuals is different at the three stations, they all show errors larger than 10 cm and a pronounced seasonal cycle. Furthermore, predominantly negative residuals indicate that Nordic4-SS overall overestimates the meteorological component. In the autocorrelation plots, the shaded areas are delimited by the confidence intervals, and values outside these bands are considered significantly autocorrelated. Note that the lags are of 12 hours because the residual time series are 12-hourly. The autocorrelation plots indicate significant non-randomness in the residuals at all three locations and, thus, a signal that has the potential to be corrected. Although the three stations have different autocorrelation patterns, they all show significant autocorrelation the first 10 days, with two longer-period peaks around two weeks and just before one month. These peaks coincide with the periods of spring and neap tides, that occur every two weeks (alternating), and the M2 tide which is on a 28 day cycle, providing evidence that the tides have an important influence on the residuals.

\subsubsection{Representation of the residuals in polar coordinates: polar plots}
\label{sec:polar_plots_results} 

In addition to this signal in time, we observe a dependence on local winds. Polar plots illustrate the variation of a magnitude in polar coordinates. 

\paragraph{Polar plots in hindcast mode}
\label{sec:polar_plots_hindcast}

\begin{figure}[ht]
  \centering
  \subfloat[Average residuals OSC]{\includegraphics[width=0.27\textwidth]{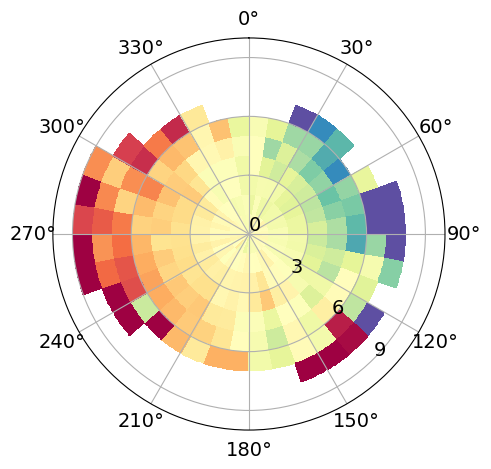}\label{fig:wind_radar_plot_mean_error_hindcast_NO_OSC}}
  \subfloat[Average residuals BGO]{\includegraphics[width=0.27\textwidth]{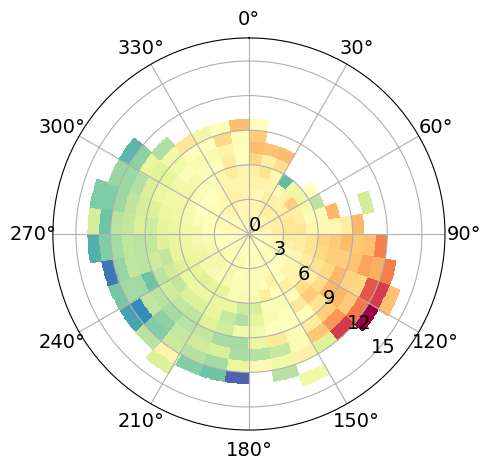}\label{fig:wind_radar_plot_mean_error_hindcast_NO_BGO}}
  \subfloat[Average residuals ANX]{\includegraphics[width=0.34\textwidth]{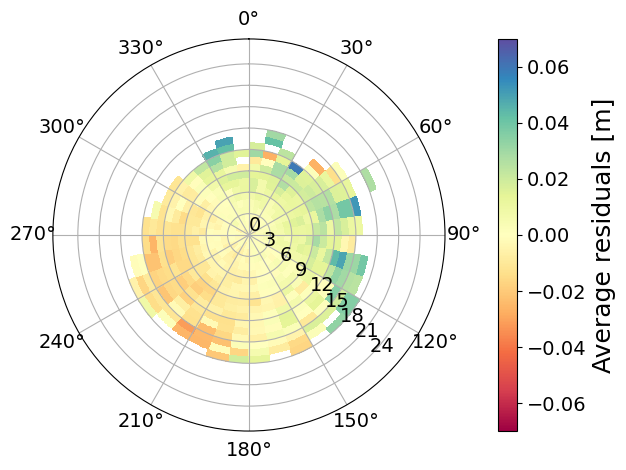}\label{fig:wind_radar_plot_mean_error_hindcast_NO_ANX}}
  
  \subfloat[Std. residuals OSC]{\includegraphics[width=0.27\textwidth]{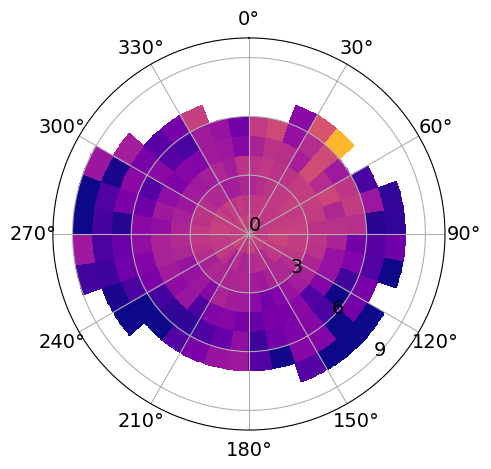}\label{fig:wind_radar_plot_std_error_hindcast_NO_OSC}}
  \subfloat[Std. residuals BGO]{\includegraphics[width=0.27\textwidth]{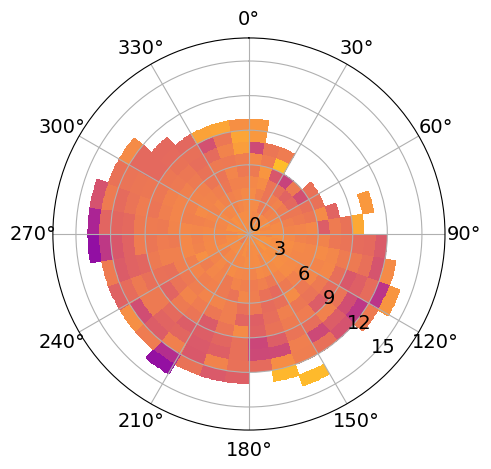}\label{fig:wind_radar_plot_std_error_hindcast_NO_BGO}}
  \subfloat[Std. residuals ANX]{\includegraphics[width=0.34\textwidth]{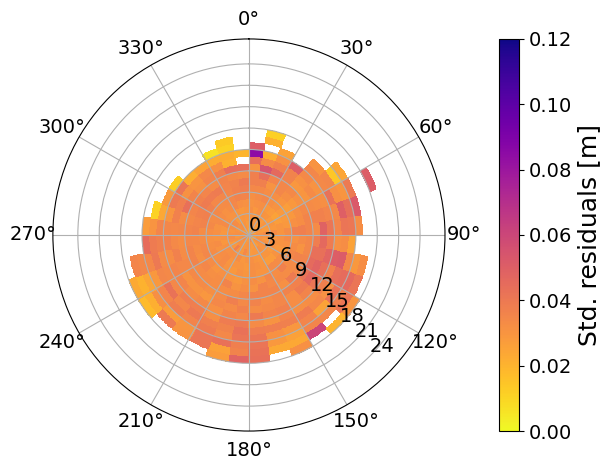}\label{fig:wind_radar_plot_std_error_hindcast_NO_ANX}}
  
  \subfloat[Nr. observations OSC]{\includegraphics[width=0.27\textwidth]{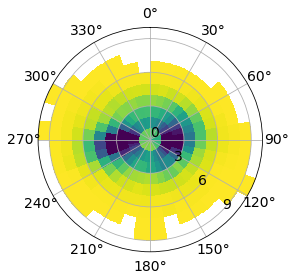}\label{fig:wind_radar_plot_count_error_hindcast_NO_OSC}}
  \subfloat[Nr. Observations BGO]{\includegraphics[width=0.27\textwidth]{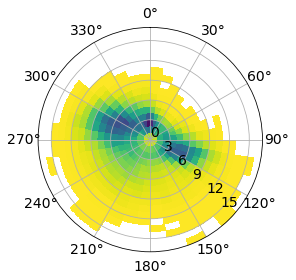}\label{fig:wind_radar_plot_count_error_hindcast_NO_BGO}}
  \subfloat[Nr. observations ANX]{\includegraphics[width=0.34\textwidth]{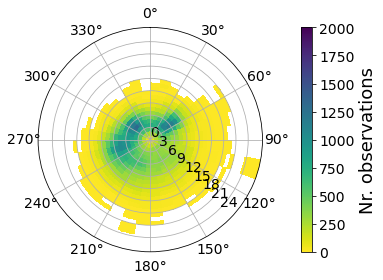}\label{fig:wind_radar_plot_count_error_hindcast_NO_ANX}}
  
  \caption{Polar plots of the statistics in \acrshort{metno}'s storm surge hindcast (NORA-SS) conditioned on wind speed and direction. The residuals have been computed as the difference between the observed and predicted meteorological component, where the predicted meteorological component is the output from NORA-SS corrected with a weighted differences correction method. The panels show the average residuals at a) \acrfull{osc}; b) \acrfull{bgo}; and c) \acrfull{anx} (red colors indicate an underestimation and blue colors indicate an overestimation by the hindcast); the standard deviation of the residuals at d) \acrshort{osc}; e) \acrshort{bgo}; and f) \acrshort{anx}; the number of observations at g) \acrfull{osc}; h) \acrshort{bgo}; and i) \acrshort{anx}. The bins are defined as boxes of size 1 m/s $\times$ 10 deg. In figures a) to f), only bins with at least five observations are colored. The figures were constructed with hourly hindcast data from the period 2000--2019. Note that the colorbar saturates for the highest values in panel a.}
  \label{fig:polar_plots_hindcast}
\end{figure}
The average and standard deviation of the residuals in the hindcast NORA-SS, as well as the total number of observations, conditioned on wind speed and direction from ERA5, are illustrated in Fig. \ref{fig:polar_plots_hindcast}. If we focus first on the radial coordinate, we can observe some similarities among the three stations, even though they are affected by different dynamics and geographical conditions. For instance, the predictions at all three locations are more accurate and exact when the wind is weak. The systematic error and the uncertainty tend to increase with wind speed in all of them, while, unfortunately, the number of observations decreases, leading to more random noise in the plots. Still, it is important to highlight that the magnitude of the wind speeds registered has large variations across the stations, and so has the bias. For instance, the bias and the standard deviation at 8 m/s are much greater at \acrshort{osc} than at \acrshort{bgo} or \acrshort{anx}. \acrshort{osc} also has a lower density of observations at 8 m/s than the other two stations, corresponding with the local climate.

The dependence of the bias on the wind direction is a local characteristic with strong spatial variability. For example, the well-defined pattern at the station \acrshort{osc} indicates that when the wind blows approximately from the north to the southeast, the model overestimates the surges; contrarily, when the wind has an eastward component, the model underestimates the \acrshort{ssh}. The pattern of the systematic error at OSC differs significantly from that at BGO and ANX. When comparing these stations, in addition to considering their geographical conditions, it is important to remember that they have different climates. Bergen is located on the West Coast of Norway, and is affected by a number of cyclones each season that push the water against the coast while the pressure is low. Oscarsborg, on the other hand, has a continental climate, and experiences calm to moderate wind conditions due to its sheltered location. Andenes is usually affected by higher tides and low-pressure systems generated near Iceland. These different conditions are well represented in our data. Wind speeds are greatest at ANX, where winds of 24 m/s have been observed, followed by BGO, where the maximum records are about 15 m/s. In contrast, at OSC, all wind records are below 9 m/s. In summary, although the patterns of the mean error at the three stations are very clear, the dependence on the wind direction is completely different. This information has been compiled in lookup tables to serve as a guide for the meteorologist on duty.

We have already mentioned that NORA-SS contains many more samples than Nordic4-SS (66536 vs 2372), so it provides more robust statistics. At the same time, the hindcast is forced with reanalysis data (NORA3) instead of the atmospheric forecasts (MEPS) used to force Nordic4-SS. Hence, the hindcast can overestimate the real forecast skill and underestimate its variability. In this sense, the results obtained for the hindcast are idealized, as it represents conditions that can not be reproduced operationally. Still, the figures shown above, generated with the NORA-SS, strongly suggesting of a pronounced local systematic error in the \acrshort{roms} model setup, even under idealized conditions. 

\paragraph{Polar plots in forecast mode}

\begin{figure}[ht]
  \centering
  \subfloat[Average residuals OSC]{\includegraphics[width=0.34\textwidth]{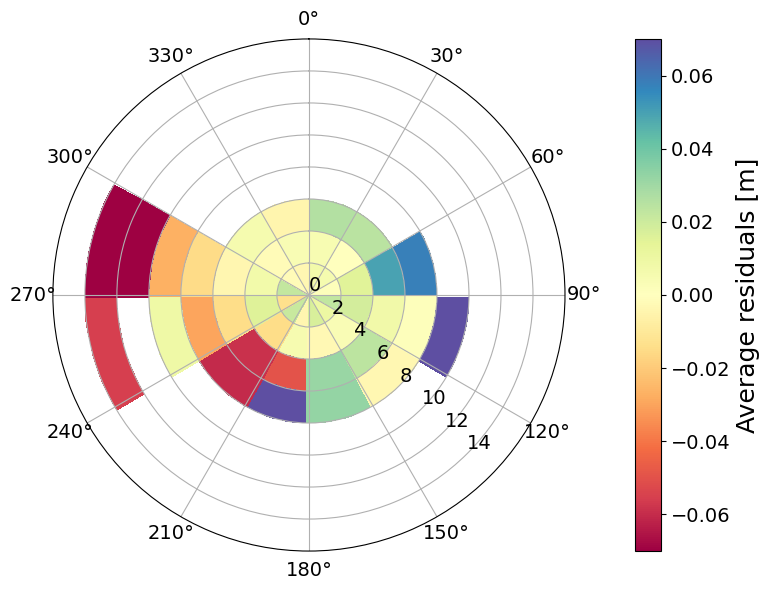}\label{fig:wind_radar_plot_mean_error_operational_NO_OSC}}
  \subfloat[Std. residuals OSC]{\includegraphics[width=0.34\textwidth]{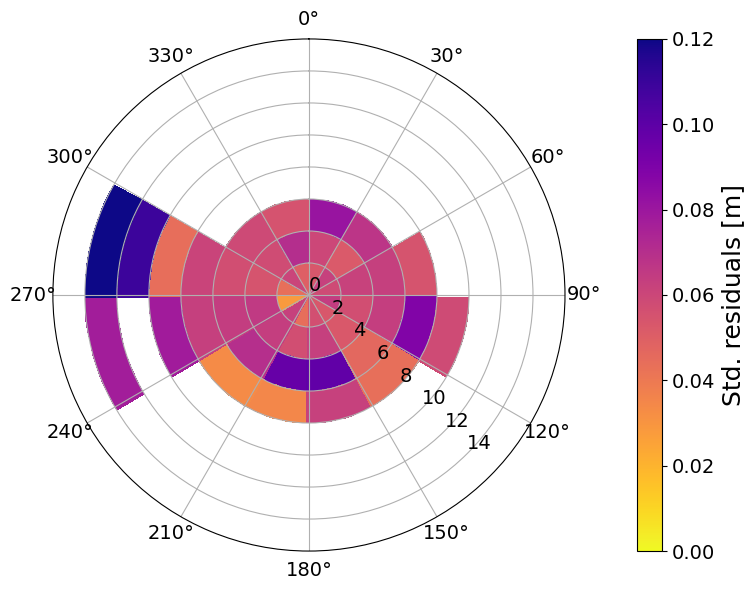}\label{fig:wind_radar_plot_std_error_operational_NO_OSC}}
  \subfloat[Nr. observations OSC]{\includegraphics[width=0.34\textwidth]{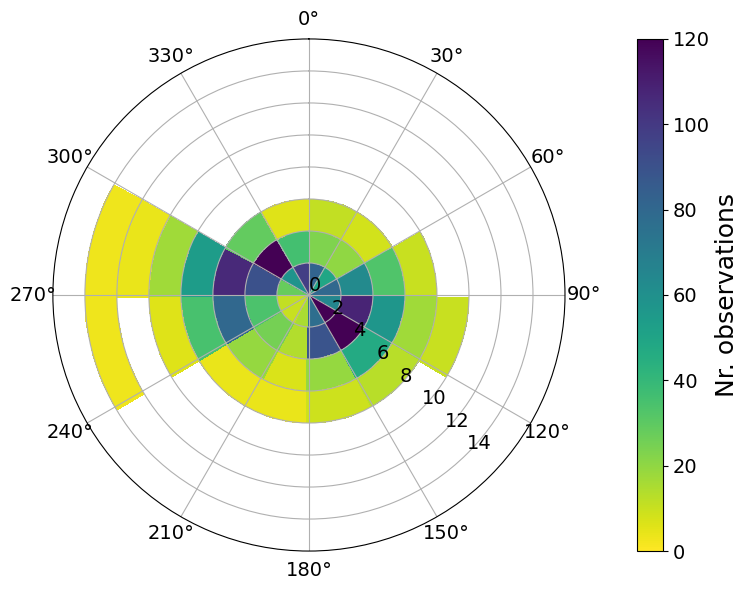}\label{fig:wind_radar_plot_count_error_operational_NO_OSC}}
  
  \caption{Polar plots of the statistics in \acrshort{metno}'s operational storm surge model (Nordic4-SS) conditioned on wind speed and direction. The residuals have been computed as the difference between the observed and predicted meteorological component, where the predicted meteorological component is the output from NORA-SS corrected with the weighted differences correction method. The panels show a) the average residuals at \acrfull{osc},  where red colors indicate an underestimation, and blue colors indicate an overestimation by the hindcast; b) the standard deviation of the residuals at \acrshort{osc}; and c) the number of observations at \acrshort{osc}. The bins are defined as boxes of size 1 m/s $\times$ 10 deg. In figures a) to f), only bins with at least three observations are colored. Note that the figures were constructed with 12-hourly 0-lead-time data from the period January 2018--March 2021. Note that the colorbars saturate for the highest values.}
  \label{fig:polar_plots_operational_no_osc}
\end{figure}

\label{sec:polar_plots_forecast}
The conditioned statistics of the residuals are also computed for the forecast data (Nordic4-SS) for the period January 2018 to March 2021. We show the average and standard deviation of the residuals, and the number of observations at \acrshort{osc} in Figures \ref{fig:wind_radar_plot_mean_error_operational_NO_OSC}, \ref{fig:wind_radar_plot_std_error_operational_NO_OSC}, and \ref{fig:wind_radar_plot_count_error_operational_NO_OSC}, respectively. Given that the forecasts are available for a much shorter period than the hindcast and that we have 12-hourly data instead of hourly data for 0-lead-time conditions, we have aggregated the results into larger bins. For the forecast data, the bins have a size of 2 m/s $\times$ 30 deg. Furthermore, we plot the values when there are at least three observations in the bin. Even though the resolution is coarser compared to the hindcast polar plots, the overall patterns agree, confirming that the detailed structure in the hindcast is not misleading despite not having the same error distributions, which corresponds to our idea that the imperfections of the \acrshort{roms} setup play a systematic role in the structure of the residuals. Moreover, these plots have a practical use in the context of the decision support system, as the coarser resolution facilitates the forecaster's decision-making. An interesting difference between the hindcast and the forecast polar is that the forecast shows much greater wind speeds. A possible explanation is that ERA5 has a coarser resolution that MEPS, and shows average values over grid cells. Also, MEPS has been developed for Scandinavia and the Nordic Seas, while ERA5 is a global dataset. A comparison of ERA5 and MEPS is beyond the scope of this study, although it would help to clarify this difference.

Similar polar plots have been generated for significant wave height in the radial coordinate and mean wave direction in the angular coordinate (examples are provided in Appendix \ref{sec:appendix_e}). We see that the dependence of the residuals with these wave parameters is in line with the results obtained for wind speed and direction. The residuals also depend on the local pressure. By binning the error at intervals of 5 hPa, we see that the numerical model overestimates the meteorological component when the pressure is low and underestimates it when the pressure is high (not shown here). This is a general result valid for all the stations. Nevertheless, it must be interpreted with caution, as the domain regions with the highest and lowest pressure values observed also have fewer observations and therefore the highest uncertainties. 

\subsection{Correction of the numerical storm surge model}
\label{sec:ml_results}

From the results shown above, it is clear that the residuals in the numerical storm surge model are correlated in time and depend on the meteorological conditions. We have tested two residual techniques: one based on polar plots and one based on \acrshort{nn}s. At the first attempt, we learned the residuals in the hindcast and tried to use this bias to correct the forecast data. Unfortunately, we then discovered that the error distribution in the hindcast and forecast data are different enough to make such transfer learning ineffective when using \acrshort{nn}s, leading to higher \acrshort{rmse}. Therefore, we focus only on correcting the operational model currently used by \acrshort{metno} (Nordic4-SS), using forecast-compatible datasets that will allow us to operationalize the correction process in the future. 

\subsubsection{Bias correction with polar plots}

The polar plots in Figs. \ref{fig:polar_plots_hindcast} and \ref{fig:polar_plots_operational_no_osc} show an evident structure in the systematic error, i.e., a dependence on both wind speed and wind direction at all locations. This is the error that we want to correct using data-driven models. Nonetheless, directly removing the bias observed in the polar plots from the Nordic4-SS forecasts does not lead to a meaningful enhancement of the model; only a few millimeters of improvement are obtained, which is less than the estimated wave gauge measurement error. We see, however, that the model's performance is sensitive to the periods chosen and the size of the bins, which we interpret as a consequence of using short forecast time series.   

Experiments conducted in hindcast mode, using longer time series from NORA-SS, show that the polar plots, in fact, have the ability to reduce the \acrshort{rmse} at some locations. For instance, at TRG, the relative improvement after removing the bias in the polar plots is of $4\%$. Meanwhile, at OSL and VIK the improvement is of $3\%$. In spite of that, the method is inefficient when extreme weather occurs and uncertainties are high. This is because the method consists of subtracting the mean bias computed for each bin and, even in the hindcast, there are not enough situations represented in the bins that correspond to severe storms. 

\subsubsection{Residual correction with Neural Networks}

We model the nonlinearities in Nordic4-SS residuals with \acrshort{nn}s, training one model (same architecture but different predictors) for each station in mainland Norway, and we run it for every lead time from $t+1$ to $t+60$. Then, we subtract the learned residuals from Nordic4SS. Although the residuals also depend on wave and pressure conditions, experiments indicate that wind speed and direction are the most important predictors and that providing wave and pressure data in addition to wind data to the \acrshort{nn}s does not improve the results. For this reason, the \acrshort{ml} algorithms presented here are trained with wind data in addition to storm surge predictions, tide data, and past \acrshort{ssh} observations. 

 \begin{figure}[ht]
  \centering
  \subfloat[RMSE residuals OSC]{\includegraphics[width=0.3\textwidth]{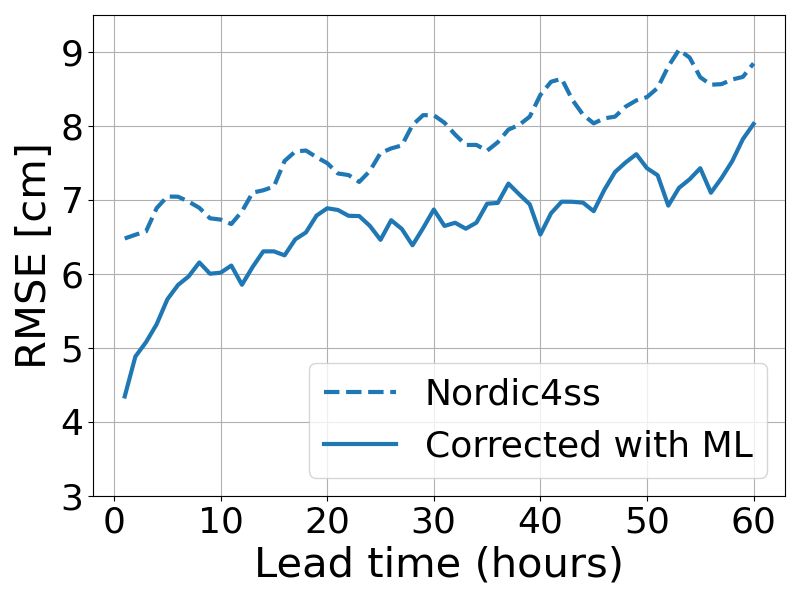}\label{fig:rmse_NO_OSC}}
  \hfill
  \subfloat[RMSE residuals BGO]{\includegraphics[width=0.3\textwidth]{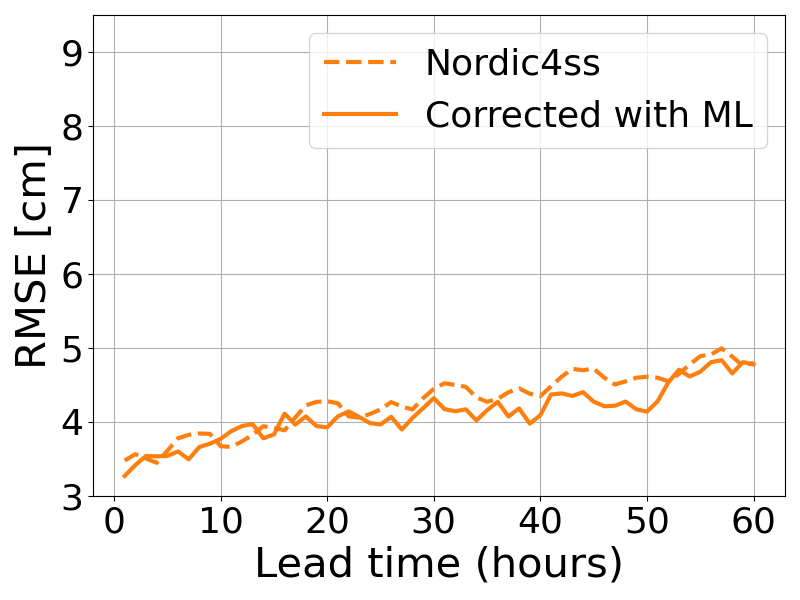}\label{fig:rmse_NO_BGO}}
  \hfill
  \subfloat[RMSE residuals ANX]{\includegraphics[width=0.3\textwidth]{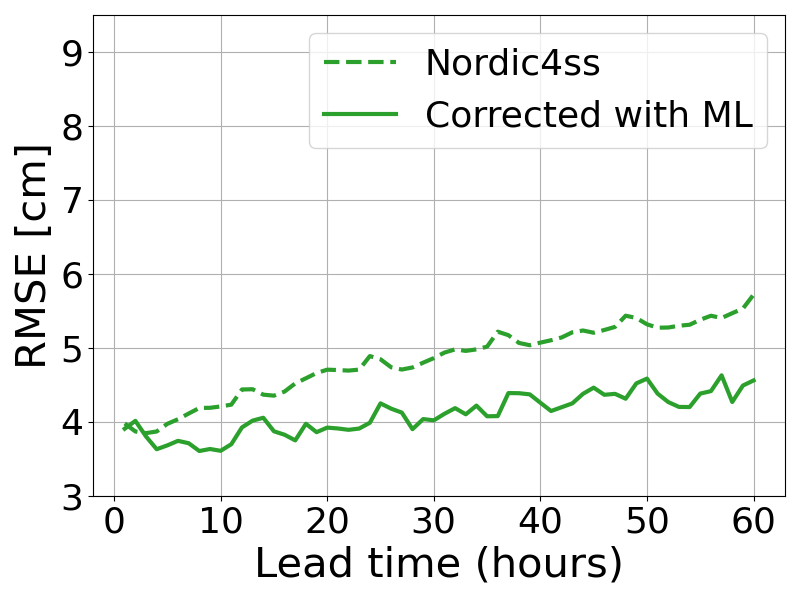}\label{fig:rmse_NO_ANX}}
  
  \subfloat[Bias residuals OSC]{\includegraphics[width=0.3\textwidth]{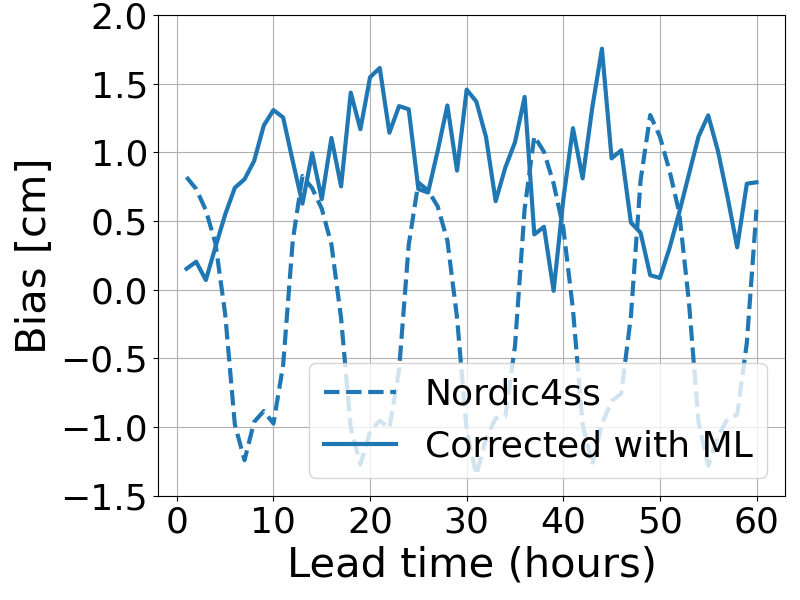}\label{fig:bias_NO_OSC}}
  \hfill
  \subfloat[Bias residuals BGO]{\includegraphics[width=0.3\textwidth]{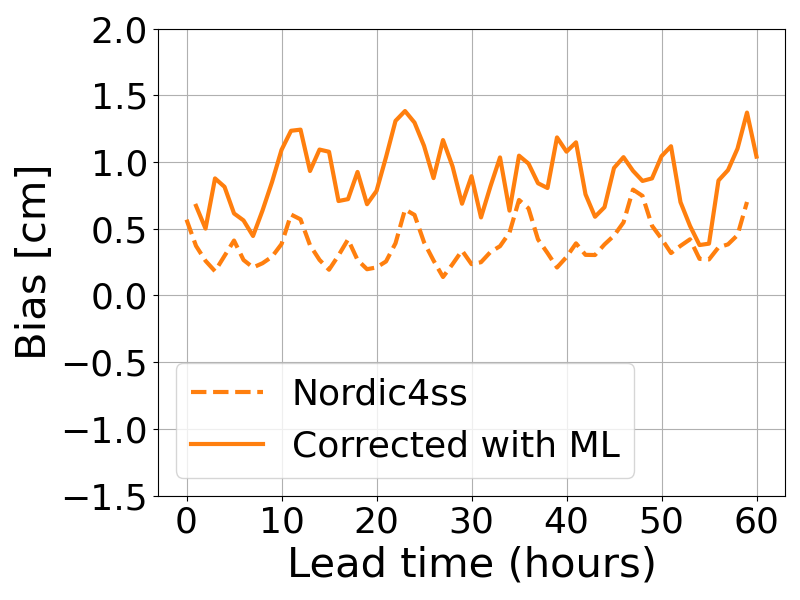}\label{fig:bias_NO_BGO}}
  \hfill
  \subfloat[Bias residuals ANX]{\includegraphics[width=0.3\textwidth]{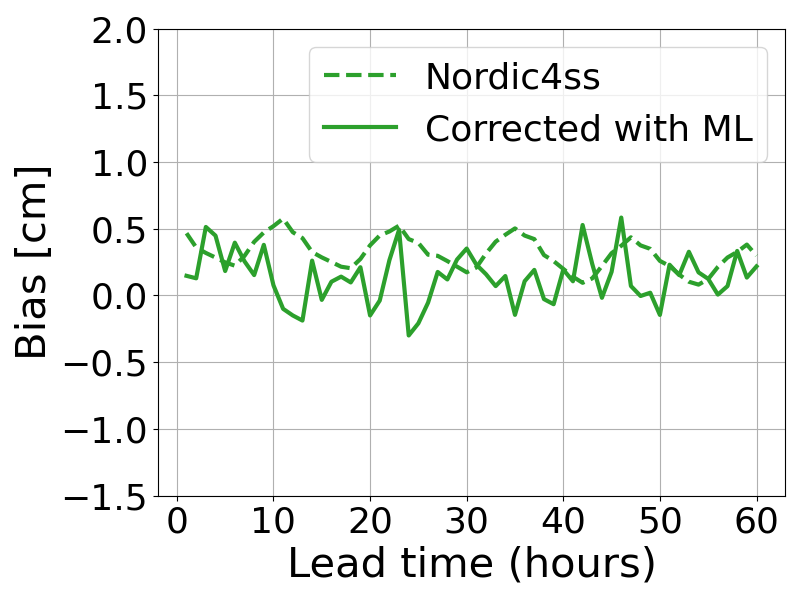}\label{fig:std_NO_ANX}}

  \subfloat[Std. residuals OSC]{\includegraphics[width=0.3\textwidth]{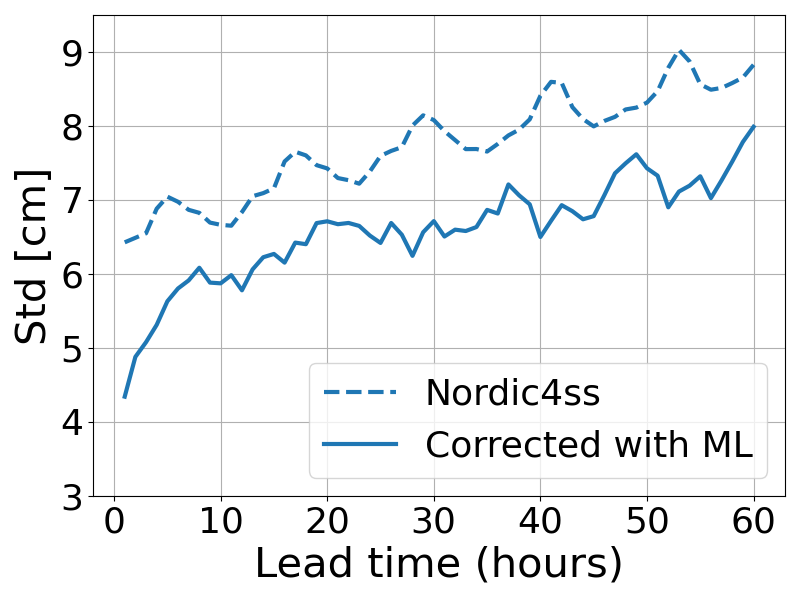}\label{fig:std_NO_OSC}}
  \hfill
  \subfloat[Std. residuals BGO]{\includegraphics[width=0.3\textwidth]{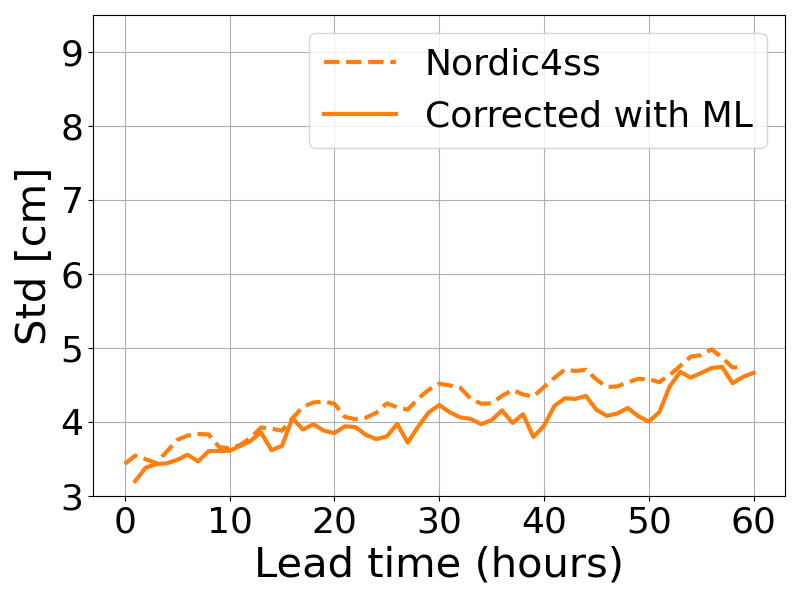}\label{fig:std_NO_BGO}}
  \hfill
  \subfloat[Std. residuals ANX]{\includegraphics[width=0.3\textwidth]{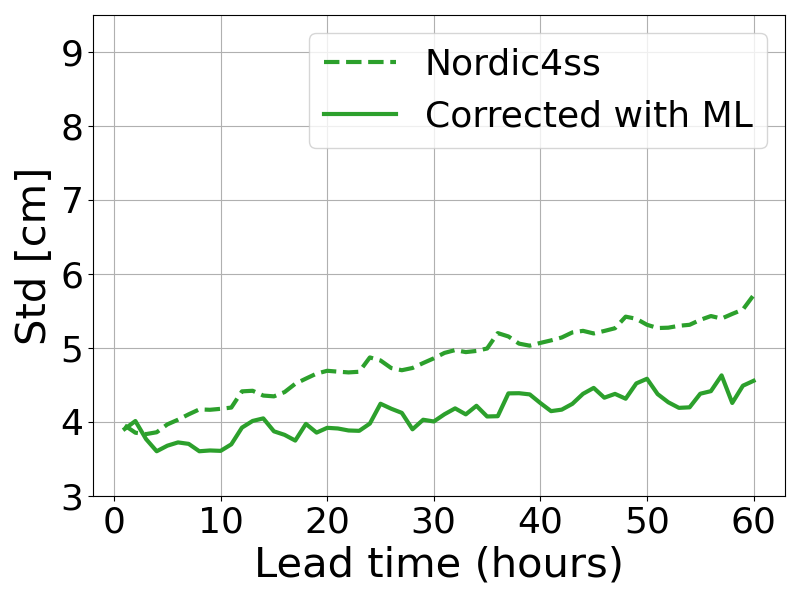}\label{fig:bias_NO_ANX}}

  \caption{Statistics of the residuals in the operational storm surge model (Nordic4-SS), before (dashed line) and after (solid line) the \acrshort{ml} correction. The residuals have been computed as the difference between the observed and predicted meteorological component, where the predicted meteorological component is the output from Nordic4-SS corrected with a weighted differences correction method. The panels show the RMSE of the residuals at a) \acrfull{osc}; b) \acrfull{bgo}; and c) \acrfull{anx}; the bias in the residuals at d) \acrshort{osc}; e) \acrshort{bgo}; and f) \acrshort{anx}; the standard deviation of the residuals at g) \acrshort{osc}; h) \acrshort{bgo}; and i) \acrshort{anx}. Note that the figures were constructed with 12-hourly data from January 2018--March 2021.}
  \label{fig:best_mode_stats}
\end{figure}

 Fig.  \ref{fig:best_mode_stats} exposes the spatial variability in the \acrshort{rmse}, bias, and standard deviation of the residuals as a function of lead time from $t+1$ to $t+60$. The figure shows Nordic4-SS forecasts (dashed lines) and the \acrshort{nn}-corrected forecast (solid lines) at three locations: \acrshort{osc} (blue), \acrshort{bgo} (orange), and \acrshort{anx} (green). The \acrshort {rmse} in the operational storm surge data typically increases with lead time at all stations, but the steepness of the curve depends on the station considered (Figs. \ref{fig:rmse_NO_OSC}, \ref{fig:rmse_NO_BGO}, and \ref{fig:rmse_NO_ANX}). The \acrshort{rmse} at \acrshort{osc} is reduced for all lead times by approximately 1 cm. If we decompose the \acrshort{rmse} into the bias and standard deviation, we see that the bias is the smallest component, and it is mostly positive after correcting with \acrshort{nn}s. It is unclear why the bias oscillates, but we can see that after correcting with \acrshort{nn}s  these oscillations are out of phase with the bias in Nordic4-SS. If we look at the results from \acrshort{bgo}, we see that the performance of the \acrshort{nn}s is poor and almost does not provide any improvement, at least for the first 24 hours. Remember that \acrshort{bgo} is one of the stations located further west in Norway, and that cyclones often move to the east or northeast. For this reason, we interpret this result as a lack of metocean information from remote western locations, affecting the \acrshort{nn}s capacity to learn and anticipate the atmospheric conditions at \acrshort{bgo}. The bias at \acrshort{bgo} is also increased after applying the post-processing method. Turning now to the results at \acrshort{anx}, we observe that the \acrshort{nn}s improve the results and that this improvement increases with lead time, from 0.5 cm at $t+10$ to 1 cm at $t+60$. The bias is also reduced for almost all the lead times. Moreover, 12-hours oscillations are observed in the curves in Fig.  \ref{fig:best_mode_stats}. These oscillations can be attributed to a contamination of the tidal component \citep{kristensen_et_al_2022} . 

\paragraph{Spatial distribution of the residuals in the numerical model and relative improvement}
\label{spatial_distribution}

\begin{figure*}[ht]
  \centering
  \subfloat[RMSE residuals at $t+1$]{\includegraphics[width=0.3\linewidth]{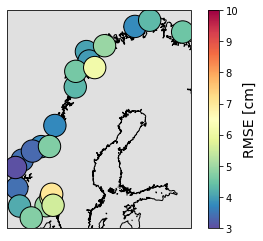}\label{fig:rmse_map_1}}
  \hspace{1em}
  \subfloat[RMSE residuals at  $t+24$]{\includegraphics[width=0.3\linewidth]{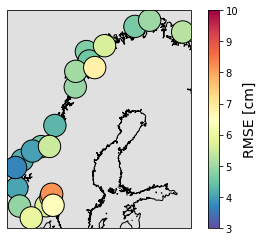}\label{fig:rmse_map_24}}
  \hspace{1em}
   \subfloat[RMSE residuals at  $t+48$]{\includegraphics[width=0.3\linewidth]{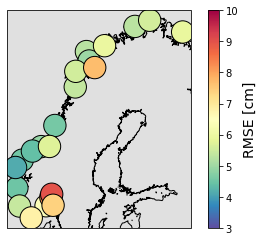}\label{fig:rmse_map_48}}
  
   \subfloat[Improvement at  $t+1$]{\includegraphics[width=0.3\linewidth]{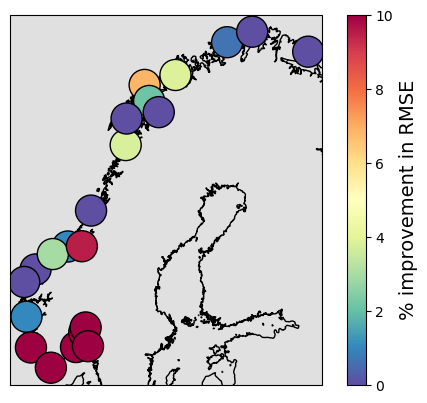}\label{perc_change_1}}
  \hspace{1em}
  \subfloat[Improvement at  $t+24$]{\includegraphics[width=0.3\linewidth]{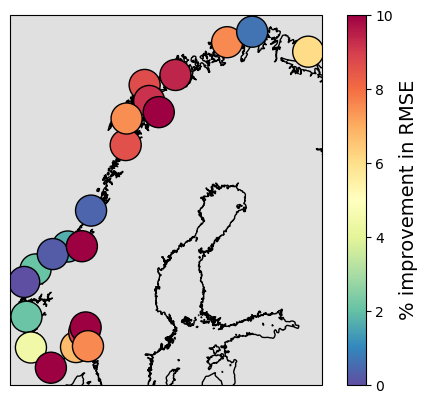}\label{fig:perc_change_24}}
  \hspace{1em}
   \subfloat[Improvement at  $t+48$]{\includegraphics[width=0.3\linewidth]{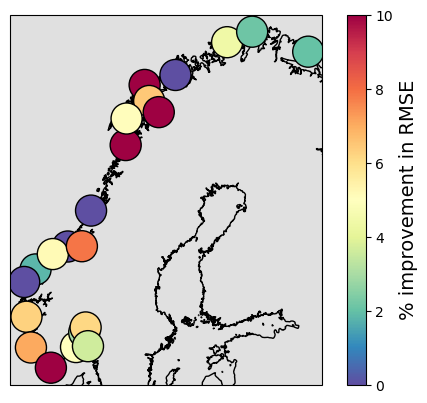}\label{fig:perc_change_48}}

  \caption{The first row of panels shows RMSE of the residuals in \acrshort{metno}'s operational storm surge model, Nordic4-SS, for lead time a) 1 hour, b) 24 hours, and c) 48 hours at all the 22 permanent harbors in mainland Norway. The second row of panels shows improvement in \acrshort{rmse} after applying the residual \acrshort{nn} correction at a lead time of d) 1 hours, e) 24 hours, and f) 48 hours. The RMSE has been computed using only one year of data, from April 2020--March 2021. Note that this is a very short period consisting of only 631 records and that the RMSE is sensitive to the period chosen.}\label{fig:change_rmse_ml}
  
\end{figure*}

The \acrshort{nn} residual method can be applied at any location where both predictions from Nordic4-SS and water level measurements are available. The first row of Fig.  \ref{fig:change_rmse_ml} shows the \acrshort{rmse} of the Nordic4-SS forecasts at all the 22 stations for $t+1$, $t+24$, and $t+48$. We see that the \acrshort{rmse} is lowest on the West Coast, and highest in Skagerrak and in the inner parts of the fjords. Furthermore, it increases with lead time. The second row of Fig. \ref{fig:change_rmse_ml} shows the percentage improvement in \acrshort{rmse} after correcting Nordic4-SS with \acrshort{nn} for lead times $t+1$, $t+24$, $t+48$. At $t+1$, the stations in Skagerrak show most improvement ($36\%$ at \acrshort{osc}). As lead time increases, the \acrshort{nn}s show a better performance in Northern Norway. Note that HVG and VAW have different characteristics than the other stations in Northern Norway, which is reflected in the results. The poorest performance of the \acrshort{nn}s is observed in Western Norway, where the storm surge values and the error in Nordic4-SS are lowest.

\paragraph{Application of the ML correction method to a storm surge event}
\label{application}

We have selected two storm surge events to illustrate the improvements in the surge predictions in forecast mode after applying the \acrshort{ml} correction for a lead time of one hour. From a historical perspective, these events might not be the most extreme, but they were the only two events in \acrshort{osc} that registered storm surges above 60 cm in the short test period (April 2020 to March 2021). Still, they provide an indication of the performance of the models when \acrshort{ssh} values are anomalous high.

\begin{figure*}[ht]
  \centering
\subfloat[Storm surge at t+1]{\includegraphics[scale=0.4]{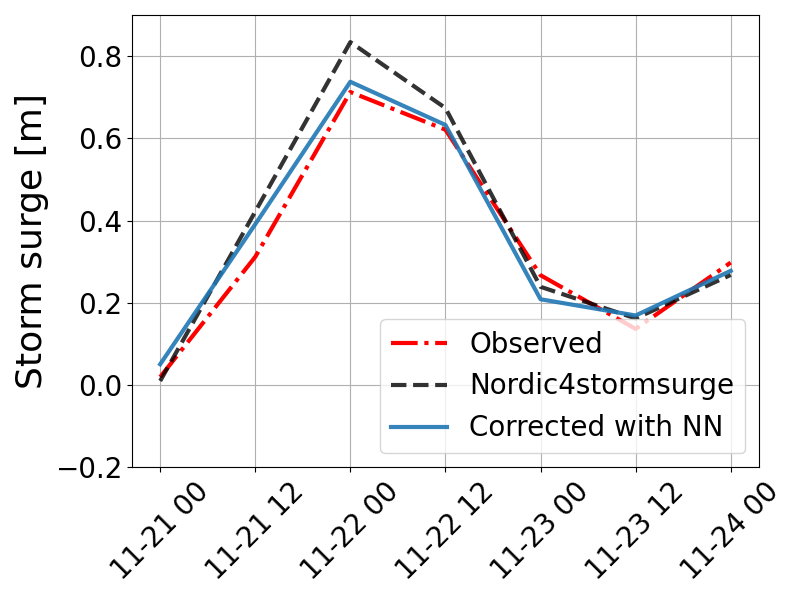}\label{fig:case_ml_oper_stormsurge_nov2020}}
  \hfill
  \subfloat[Residuals at t+1]{\includegraphics[scale=0.4]{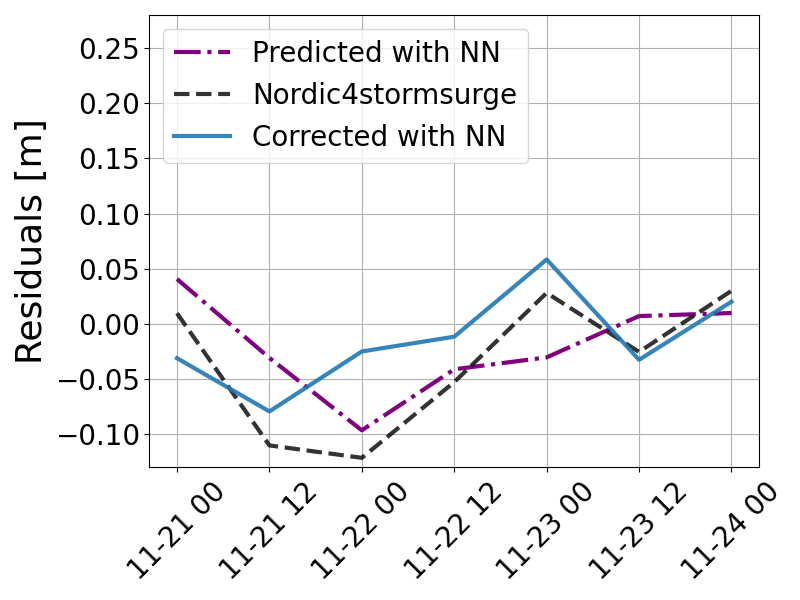}\label{fig:case_ml_oper_stormsurge_nov2020_error}}

  \subfloat[Storm surge at t+1]{\includegraphics[scale=0.4]{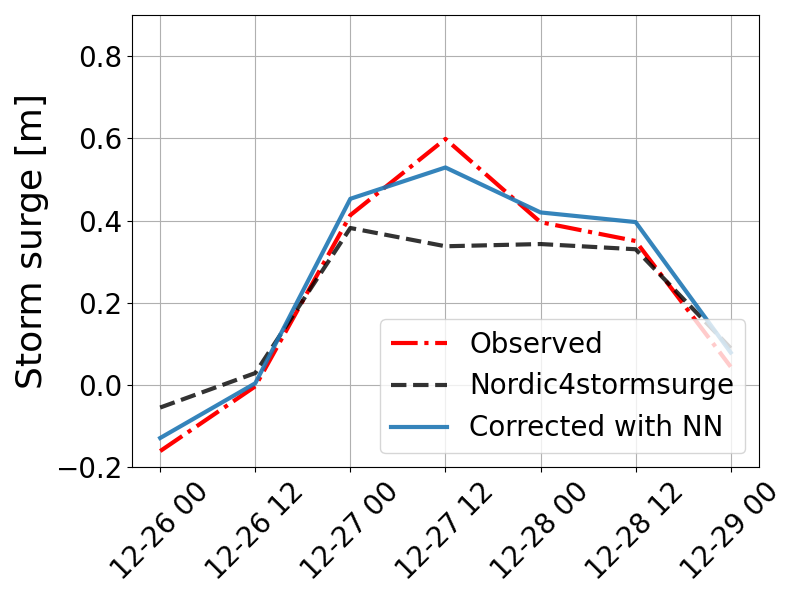}\label{fig:case_ml_oper_stormsurge_dec2020}}
  \hfill
  \subfloat[Residuals at t+1]{\includegraphics[scale=0.4]{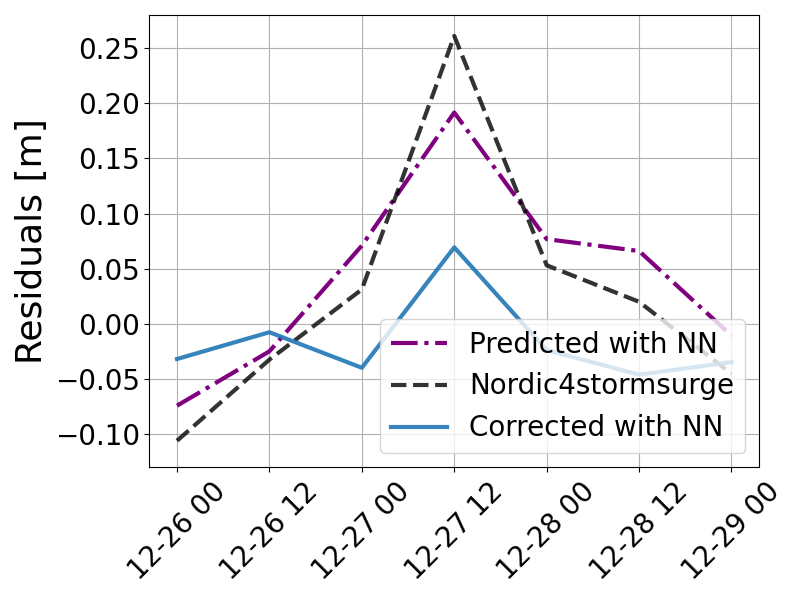}\label{fig:case_ml_oper_stormsurge_dec2020_error}}

  \caption{Storm surge and residuals of two events at the Norwegian harbor \acrfull{osc}. Panel a) shows the observed storm surge with a dashed-dotted red line, the predicted storm surge from Nordic4-SS with a dashed black line, and the storm surge from Nordic4-SS corrected with \acrshort{nn}s with a solid blue line, for the event on November 21--23, 2020. Panel b) shows the residuals in Nordic4-SS predicted with neural networks with a dashed-dotted purple line, the residuals in Nordic4-SS with \acrshort{nn}s with a dashed black line, and the residuals from Nordic4-SS after correcting with \acrshort{nn}s with a solid blue line, for the event on November 21--23, 2020. Panels c) and d) are equivalent to a) and b), but for the event on December 27--28, 2020.}
  \label{fig:cases_2020}
\end{figure*}

The first event was registered on November 21--23, and was associated with a low pressure system that originated south of Iceland and traveled to Northern Norway, causing strong westerly winds in Southern Norway. Due to this event, \acrshort{metno} had to issue warnings of high water levels (with an uncertainty of 5--10 cm) and strong winds. The consequences associated with the sea-level warning were local floods and risk of small damage on infrastructure and buildings along the coastline. The event is illustrated in Fig.\ref{fig:case_ml_oper_stormsurge_nov2020}. This figure was constructed by taking the values of lead time $t+1$ associated with seven forecast runs. Note that each Nordic4-SS forecast and, subsequently, each \acrshort{ml} prediction is generated every 12 hours. We see that the predicted storm surge was above 80 cm, overestimating the actual event that resulted in a storm surge of about 70 cm. At the peak of the event, the error in Nordic4-SS is -12 cm (Fig. \ref{fig:case_ml_oper_stormsurge_nov2020_error}), slightly outside the uncertainty range, but after correcting with \acrshort{nn}s the error is reduced to 4 cm. Moreover, the wind speed and direction predicted with MEPS for the peak of the event were of 4 m/s  and $260 ^\circ$. The average residuals in the polar plots for these values indicate a moderate underestimation of Nordic4-SS, which is the opposite of what occurred.

Figs.  \ref{fig:case_ml_oper_stormsurge_dec2020} and \ref{fig:case_ml_oper_stormsurge_dec2020_error} show the storm surge and residuals, respectively, of the event caused by storm Bella on December 26--27, 2020. A large area in the North Atlantic was affected by a low-pressure system that originated when an intrusion of cold air from Canada met relatively warm air over the Western Atlantic. The Jet Stream helped develop and deepen the low-pressure system before it continued its track to the North Sea. While interacting with the Azores high, it generated strong winds and heavy rains. The storm had an enormous impact across the Norwegian territory, forcing \acrshort{metno} to issue 62 warnings because of strong winds, floods, and land- and snow slides. We can see that the contribution of the storm surge to the rise in \acrshort{ssh} was of 60 cm on December 27 at 00 hours at \acrshort{osc}. Nordic4-SS predicted the rise in \acrshort{ssh} associated with the storm, but the fall was predicted a couple of hours before it actually occurred. Consequently, the numerical model underestimated the maximum levels by more than 20 cm. On the other hand, we see that the residual learning method can detect the delay, reducing the error by more than 10 cm. Unfortunately, for longer lead times, the \acrshort{nn}s do not perform equally well for this extreme event. This shows that our correction technique can be used for short term updates to the predictions issued at the reference stations. In addition, this is an event where the forecasters could have used the polar plots to adjust the forecasts. The wind speed and direction forecasted with MEPS for the peak of the event at \acrshort{osc} were 12 m/s and $284 ^\circ$ respectively. For these values, the polar plots indicate that Nordic4-SS typically strongly underestimates storm surges, which was the case.

\section{Summary and Discussion}
\label{sec:summary_discussion}


The aim of this study is to improve Nordic4-SS, the numerical storm surge model that runs at \acrshort{metno}. First, we show that the model has systematic errors that depend on atmospheric conditions and the geographical location of the station considered. As part of the validation process, we represent the dependence of the error in the model relative to the local wind speed and direction in polar coordinates. We also find that the residuals are significantly autocorrelated. Put together, these results indicate that the residuals in the numerical model are not random and that there is a structure that could be learned using data-driven models. 

Given that the numerical model already has a good performance, that decades of development have been dedicated to understanding the physical processes, that it is generally very reliable, and that the meteorologists on duty are trained to use this model, we consider it most adequate to reduce the errors in this model using a post-processing approach, rather than training a new data-driven model to compete with Nordic4-SS. From a practical point of view, the methods proposed are particularly convenient because they run on top of \acrshort{metno}'s model, meaning that they can be readily used in practice and that it is possible to compare the current predictions to the corrected values. Another advantage of this post-processing method is that the forecasters need no knowledge of \acrshort{ml} to apply the result. Furthermore, although correcting a physical model with a coarse resolution can reduce the computational cost of running a numerical model \citep{pasquali_2020}, our experience from this study indicates that it is also computationally possible to run a physical storm surge model with the required resolution for operational purposes and make the \acrshort{ml} correction on top of it to improve its quality. The additional computational time spent on correcting Nordic4-SS with the \acrshort{nn}s is in the order of seconds per station.

When computing the polar plots, we observe that the error and the uncertainty tend to increase with wind speed at all locations. The fact that the error in the storm surge model is big when winds are strong, has naturally strong implications for the prediction of extreme surge events, which are the ones we are most concerned about because they can cause the most damage. The dependence of the bias on wind direction, on the other hand, is a characteristic of each station. Overall, the patterns observed in the polar plots are in line with the experience-based intuition of the forecasters and help build confidence in our correction methods. We explored the possibility of using polar plots to correct the operational numerical model by removing the bias conditioned on local wind speed and direction from the storm surge forecasts. However, we were not able to obtain a meaningful enhancement of the forecasts, and only about $3-4 \%$ improvement at some locations for the hindcast. The strengths of this method rely on the fact that it is possible to gain physical understanding of the local bias. A drawback is that we cannot simultaneously apply the method to correlated variables, which means that we cannot subtract the bias computed for different variables because we would end up removing the correlated part of the bias twice. The poor performance of the method, and the fact that we cannot correct the error associated with multiple variables, along with the nonlinear nature of the problem, motivates the use of \acrshort{nn}s for bias correction of the numerical storm surge model.
  
Thus, the second method developed for improving the numerical storm surge model consists of learning the residuals at each Norwegian water level measurement station with \acrshort{nn} models. With this method, we managed to improve the skill in Nordic4-SS for lead times up to 60 hours at several of the permanent Norwegian stations, in particular those located in Skagerrak and Northern Norway. The models have been validated by comparing the \acrshort{rmse} of the current forecasts based on \acrshort{roms} against the \acrshort{rmse} of the corrected values. For instance, the percentage improvement at \acrshort{osc} is of 36\% for $t+1$ and 9\% for $t+24$. Although the \acrshort{nn}s outperform the polar plot method, it must be taken into account that \acrshort{nn}s are complex nonlinear methods that involve numerous computations and, as such, are harder to interpret. This might cause hesitation about applying the methods operationally. However, we believe that using this as a post-processing method and allowing the forecasters to compare the residuals estimated with \acrshort{nn}s to the polar plots, will make the method more likely to be adopted operationally, at least as an aid to the forecasters in producing their analysis.

To illustrate the applicability of the \acrshort{nn} correction method, we analyzed the forecast of the storm surge events above 60 cm in the test dataset. We found two events, one where Nordic4-SS overestimates and one where it underestimates the water levels. A comparison of the Nordic4-SS forecast and the corrected values for the event caused by storm Bella showed a reduction in the error of about $25\%$, or 10 cm. Although the improvement was computed for $t+1$, it demonstrates the benefits of applying the residual \acrshort{nn} method for predicting storm surge events. It is important to remember that the model was trained in forecast mode with only two years of data.

Despite the ability of the \acrshort{nn} to correct the storm surge predictions, there were some unexpected constraints in the correction process. First, we learned the relevance of using datasets that are continuous in time and rely on the exact same model setup for the whole period. For instance, we could not transfer learning using \acrshort{nn}s from the hindcast to the operational model, as the original intention was, because of slight differences in the datasets. This was a problem even though both datasets were generated with the same modeling system, \acrshort{roms}, with the same setup and bathymetry and parameterizations. A possible explanation for why we did not succeed in transferring learning is that the output from \acrshort{roms} relies on the atmospheric forcing and how the model is initialized. The hindcast is initialized once a year, and uses all the data available a posteriori over the whole year to be optimized. In contrast, the operational model is initialized every 12 hours and, naturally, uses only data that are already available by the time the forecast is run. Since we were not able to apply transfer learning from the NORA-SS to Nordic4-SS, we had to split an already short forecast sample to train and test the \acrshort{nn}s. In other words, we had much less data to correct Nordic4-SS than we initially thought. It must also be mentioned that enough data is essential not only for the training process but for testing the results, because the \acrshort{rmse} is sensitive to the number of events in the test sample, which varies from year to year.

All the experiments conducted were based on the residual learning framework, which consists of learning the deviation of the predicted storm surge values from the observed values, instead of estimating the actual \acrshort{ssh}s as previous studies do \citep[e.g., ][]{bruneau_et_al_2020, tiggeloven_et_la_2021, bajo_2010, cox_et_al_2002}. As such, the targets are the residuals, where the \acrshort{roms} output has been corrected using a weighted differences correction method prior to the computation of the error. The residual learning technique has been applied for short-term predictions before \citep[e.g. ][]{cox_et_al_2002}. However, comparing the results with previous findings in the literature is not straightforward because 1) not all studies use the \acrshort{rmse} to evaluate the prediction skill of their models, 2) even when the same metric is used, we have seen that the \acrshort{rmse} and the relative improvement for data corrected with the same method have a strong spatial variability, 3) when data are scarce, the \acrshort{rmse} is also sensitive to the period chosen to train and test the models, and 4) a reduced number of studies model storm surges for lead times longer than a day. For this reason, we assessed the performance of the \acrshort{ml} models by comparing the corrected data to Nordic4-SS predictions without \acrshort{ml} correction.

We want to emphasize that data-driven methods are highly sensitive to the number of samples provided. Hence, we expect an improvement of the results when training the algorithms with longer time series. Further improvements could involve fine-tuning the parameters in the \acrshort{nn}s. For example, the polar plots illustrate how the bias depends on the location of the station. This implies that we need to train one \acrshort{nn} for each station. In this work, however, the predictor variables were chosen based on experiments conducted to improve the residuals at \acrshort{osc} at $t+1$, and the stations these predictors were selected from are determined for each of the three Norwegian regions, not individual stations. We believe that a natural progression of this work would be to explore the advantages of selecting a different set of stations and predictor variables to correct the residuals at each location and lead time. Furthermore, we could optimize the architecture of the \acrshort{nn}s for each location and lead time instead of running the same models. If we had more samples, we could also explore adding more variables, for instance, weather forecasts generated in the past, or more lags, without the risk of overfitting. We could also experiment with using inputs from a different weather model, with longer lead times than MEPS, which could allow us to correct the complete storm surge forecasts until $t+120$. Finally, it would be interesting to study extreme events and nonlinear interactions of the residuals with tides when winds are strong, and when tides are high.

To summarize, this study has shown that applying a data-driven methodology for reducing the residuals in Nordic4-SS will positively impact the efficiency of warning systems and the response to storm surge events at many Norwegian stations. The methods developed are particularly convenient because they can, with minimal cost and effort, be adopted as a post-processing tool of the operational storm surge model without changing the current procedures or setup of \acrshort{roms}. Given that the \acrshort{ml} models are trained on data that are available when making the predictions, the computational demand is manageable, and as the parameters are already optimized, they can efficiently run on top of \acrshort{metno}'s predictions. Moreover, this study sets a precedent for successful bias correction with \acrshort{nn} residual learning of an operative storm surge model and describes a simple methodology that can be applied to any numerical model, also at global scale. This data-driven method, which can be seen as a post-processing method, does not interfere with the process of improvement of the numerical model itself, but has the advantage that it is faster to develop. Therefore, it can be seen as a flexible solution for correcting errors related to the numerical models' missing physics, setup, forcing, etc. The performance of the \acrshort{nn} method is limited by the amount of observations, but as long as the numerical model is not perfect, the present study indicates that it would be beneficial to apply this correction technique.

\section*{Acknowledgements}

The authors would like to thank the Norwegian Mapping Authority for their assistance with the tide data.

This work was funded by the Machine Ocean project, grant number 303411, and the Stormrisk project, grant number 300608 ({\O}B and OJA), awarded by the Research Council of Norway.

\appendix

\section{Open source data and code release}
\label{sec:appendix_a}

All the results in this study can be reproduced with the datasets and Python code in the Github repository \url{https://github.com/paulina-t/bias_correction_of_storm_surge_model_with_nn}. This includes NetCDF files with in situ observations, tide estimates, Nordic4-SS and MEPS forecasts. Jupyter notebooks with examples and additional figures are also available in the repository.

\section{Station location}
\label{sec:appendix_b}

Nordic4-SS produces storm surge forecasts for 23 permanent Norwegian stations. We estimate the residuals at 22 of these stations and group them into three regions. Table \ref{tab:station_info} contains geographical information of all the stations. 


\begin{longtable}{|c|c|c|c|c|c| }
\hline
\toprule
StationID                   & Name                      & Region              & Latitude & Longitude \\ \midrule

AES                     & Aalesund                  & West Coast               & 62.47    & 6.15      \\
ANX                     & Andenes                   & Northern Norway          & 69.33    & 16.13     \\
BGO                     & Bergen                    & West Coast               & 60.4     & 5.32      \\
BOO                     & Bodoe                     & Northern Norway          & 67.29    & 14.4      \\
HAR                     & Harstad                   & Northern Norway          & 68.8     & 16.55     \\
HEI                     & Heimsjoe                  & West Coast               & 63.43    & 9.1       \\
HFT                     & Hammerfest                & Northern Norway          & 70.66    & 23.68     \\
HRO                     & Helgeroa                  & Skagerrak                & 59.0     & 9.86      \\
HVG                     & Honningsvaag              & Northern Norway          & 70.98    & 25.97     \\
KAB                     & Kabelvaag                 & Northern Norway          & 68.21    & 14.48     \\
KSU                     & Kristiansund              & West Coast                & 63.11    & 7.73      \\
MAY                     & Maaloey                   & West Coast               & 61.93    & 5.11      \\
NVK                     & Narvik                    & Northern Norway          & 68.43    & 17.43     \\
OSC                     & Oscarsborg                & Skagerrak                & 59.68    & 10.6      \\
OSL                     & Oslo                      & Skagerrak                & 59.91    & 10.73     \\
RVK                     & Roervik                   & West Coast               & 64.86    & 11.23     \\
SVG                     & Stavanger                 & West Coast               & 58.97    & 5.73      \\
TOS                     & Tromsoe                   & Northern Norway          & 69.65    & 18.95     \\
TRD                     & Trondheim                 & Northern Norway          & 63.44    & 10.39     \\
TRG                     & Tregde                    & Skagerrak                & 58.01    & 7.55      \\
VAW                     & Vardoe                    & Northern Norway          & 70.37    & 31.1      \\
VIK                     & Viker                     & Skagerrak                & 59.04    & 10.95     \\
 \bottomrule
\caption{Table with the location of the 22 Norwegian water level stations where Nordic4-SS has been validated and predictions have been improved with \acrshort{ml}. The columns in this table represent the station ID, full name, region, latitude ($^\circ$N), and longitude ($^\circ$W).}
\label{tab:station_info}

\end{longtable}


\section{Storm surge theory}
\label{sec:appendix_c}

This appendix briefly introduces basic storm surge theory that can help understand the results.

Storm surges are measured as the height of water above the normal predicted tide (see Fig. \ref{fig:surge_def}). Tides are mainly caused by astronomical forces and lead to the regular ebb and flow of the sea. As such, the sea-level fluctuations tides produce are highly predictable \citep{pugh_1987, haigh_2017}. In classical tidal harmonic analysis, the assumption is that tidal variations can be represented as a finite sum of a series of sines and cosines of frequencies that are multiples of the fundamental frequency \citep{pugh_1987}. At a given time $t$, such terms have the form:

\begin{equation}
    S_k \cos{(\omega_k t - G_k)},
\end{equation}

where $S_k$ is the amplitude, $\omega_k$ is the angular speed, and $G_k$ is the phase lag on the Equilibrium Tide at Greenwich. Thus, the surface elevation at a particular location and time due to tides can be expressed as a linear sum of independent constituents added to the mean sea level as in the following expression:

\begin{equation}
    S_{ap} = S_{0}(x, y) + \sum_{k=0}^n f_{k}(t)S_{k}(x, y)
            \times \cos [\omega_{k}t + {v}_{k}(t) + u_{k}(t) - G_{k}(x,y)],
\end{equation}

where:

\begin{itemize}
    \item $n$ is the number of constituents,
    \item $S_{0}(x, y)$ is the mean sea level,
    \item $S_{k}(x, y)$ is the amplitude of the constituent of index $k$,
    \item $G_{k}(x, y)$ is the phase lag relative to Greenwich time,
    \item $\omega_{k}$ is the angular frequency of the constituent of index $k$,
    \item $v_{k}$ is the astronomical argument at time $t$,
    \item $f_{k}(t)$ is the nodal correction coefficient applied to the amplitude of the constituent of index $k$.
\end{itemize}

Tidal predictions differ from observations of \acrfull{ssh} because of the weather effects. One of the components of the sea-air interaction is the effect of atmospheric pressure on the water's surface. Atmospheric pressure and sea level have an inverse relationship, denominated \acrfull{ibe}. The \acrshort{ibe} consists of a rise of the water level in the presence of low air pressure or vice versa. However, the sea level does not change instantaneously owing to the need to move water masses and the inertia in the whole ocean system, but responds to the average change in pressure over a larger area. As a general rule, if the air pressure drops by one \acrfull{hpa}, the water level rises by one centimeter, accordingly to what hydrostatics would predict. More formally, atmospheric pressure can be transformed into an equivalent inverse barometer \acrshort{ssh}, $\eta_{ib}$, as expressed by the following equation:

\begin{equation}
    \eta_{ib} = - \frac{1}{g \rho_{water}} (p_{atm} - p_{ref}),
    \label{eq:zibe}
\end{equation}

where $g$ is the local gravity, $\rho_{water}$ is the water's density, $p_{atm}$ is the atmospheric pressure, and $p_{ref}$ is the reference atmospheric pressure usually set to 1013 hPa.

Air-sea interaction is not limited to the \acrshort{ibe}. Consider the situation in which the wind blows over the ocean. The air, which moves more rapidly than the water, produces a shear stress parallel to the sea surface, transferring energy and momentum. How the wind stress affects deeper layers depends on for how long the wind blows, the strength of the turbulent coupling between the ocean and the atmosphere, the Coriolis effect, and the stratification of the water column. The effect of winds on \acrshort{ssh} is inversely proportional to the water depth. It is, therefore, more important when the wind blows over an extended shallow region. The magnitude of the turbulent wind stress, $\tau$, is often parameterized as a function of the wind speed at a certain level, $U_h$, and the air density, $\rho$, in the following way:

\begin{equation}
    \tau = \rho C_D U_h^2,
    \label{eq:wind_stress}
\end{equation}

where $C_D$ is a dimensionless wind drag coefficient. In addition, in the presence of a boundary in a rotating system, the wind stress parallel to the shore will cause an Ekman flow perpendicular to the coast. If the Ekman flow is directed towards the coast, water will be piled up \citep{gill_1982}. Earth's rotation causes winds to move toward the right in the Northern Hemisphere, such that the largest surge will be in the right forward part of the storm in this hemisphere, due to the Coriolis effect. Meanwhile, the balance between frictional forces due to wind stress and the Coriolis force will drive surface currents at an angle to the right of the wind direction in the northern hemisphere (the exact angle will vary with the turbulent properties of the fluid but will typically range from $15-30^\circ$), the well-known Ekman current. The Ekman transport (the integral over the vertical dimension) will point $90^\circ$ to the right of the wind stress vector \citep{gill_1982}. Consequently, when the Earth's rotation bends the currents into a more perpendicular direction with respect to the shore, it can amplify the surge. When the surge arrives on the Norwegian coast, it is primarily winds from the south and west that create an excess of water along the coast \citep{kartverket}.


The conservation of momentum also involves the process of momentum transfer from wind-generated waves. Applying a linear approximation for a steady state and a sea surface slope in equilibrium with a constant wind field, the surge height $\zeta$ due to wind-driven forces can be understood with the following relation \citep[e.g., ][]{resio_et_al_2008, pugh_1987}:

\begin{equation}
    \zeta \propto \frac{\tau_s}{gh}W,
\end{equation}

where $\tau_s$ is the wind stress at the sea-air interface, $g$ is the gravitational acceleration, $h$ is the depth of the water, and $W$ is the shelf width. In this approximation, we have not considered that the depth is variable.

Furthermore, in rotating fluids, Kelvin waves are a solution to the hydrodynamic equations with vanishing meridional velocity normal to a lateral boundary \citep{gill_1982}. These waves can, in theory, exist at all frequencies, also at the time scales of storm surges. Furthermore, Kelvin waves can only move along a coast in one direction, with the coast on the right in the Northern Hemisphere. A particular characteristic of Kelvin waves is that they can be generated by surface wind forcing close to the shore, tidal forces, or reflection of other waves incident on a coast. This means that strong winds in a remote location can generate long waves that travel along the coast, causing the water to rise even though the local winds are calm. For example, in the North Sea, winds can cause a Kelvin wave that propagates anti-clockwise from eastern Britain and eventually lead to high water levels on the Norwegian coast. The propagation of Kelvin waves is slower in shallow seas. In the North Sea, which has an average depth of 50 m, a Kelvin wave that originated in England can reach Norway in about a day \citep{kantha_and_clayson_2000}.

Despite the fact that the effects of the atmosphere on the \acrshort{ssh} are mainly due to wind stress and atmospheric pressure gradients, \acrshort{ssh} departures associated with storm surges depend on a wide range of parameters. Some important factors are the storm intensity, forward speed, size, angle of approach to the coast, central pressure, as well as the shape and characteristics of coastal features \citep{noaa}. For instance, a storm surge will be greater in regions with a shallow slope of the continental shelf than in regions with a steep slope. Furthermore, a narrow shelf with relatively deep waters is associated with lower surges but higher and more powerful waves. The opposite is true for narrow shelves with shallow waters. Bays are particularly vulnerable because storm surge water is funneled in. In addition to the storm's characteristics and the topography, the \acrshort{ssh} might be affected by nonlinear effects, such as the bottom friction that removes energy from the motion, finite water depth, flow curvature, and tide-surge interactions \citep{pugh_1987}. Furthermore, the rain effect can also contribute considerably to rising sea levels in estuaries. Heavy rains can cause surface runoff which can quickly flood streams and rivers, increasing the water level in estuaries. These effects are challenging to capture accurately in numerical models and introduce systematic biases in model outputs.

As mentioned before, surges are mainly generated by wind stress and low-pressure systems. The analytical expressions in Eqns. \ref{eq:zibe} and \ref{eq:wind_stress} are helpful for the physical interpretation of the processes involved. Even so, the actual response of the sea to the weather, in the presence of irregular boundaries and variable depths, is more complex and cannot be fully described by these equations. This limitation motivates the use of advanced numerical models and \acrshort{nn}s, which are nonlinear models, to model the meteorological component of the \acrshort{ssh} variations. Moreover, minor variations in weather patterns might result in very different responses, in particular in water bodies with tendencies for resonances and oscillations \citep{pugh_1987}. In this work, we use data from stations located along the Norwegian coast to develop a post-processing correction method of Nordic4-SS. 

The numerical modeling system used in this paper, \acrshort{roms}, runs in barotropic mode. Thus, the governing equations are the shallow water equations \citep{haidvogel_et_al_2008}. If the total height of the fluid column is $h = H + \zeta$, where the $H$ is the equilibrium depth and $\zeta$ is the sea surface deviation, we can integrate the velocity between $H$ and $\zeta$ to obtain the  volume flux through a fluid column $\mathbf{U}$. Then, the shallow water equations in flux form are:

\begin{equation}
    \partial_t  \mathbf{U} + \nabla_H . (\frac{ \mathbf{UU}}{h}) + f  \mathbf{k} \times  \mathbf{U} = -gh \nabla \zeta + \rho_0 ^{-1} (\mathbf{\tau}_s - \mathbf{\tau}_b) + \mathbf{X}
    \label{eq:swe_1}
\end{equation}

and

\begin{equation}
    \partial_t  h + \nabla_H . \mathbf{U},
    \label{eq:swe_2}
\end{equation}

where  $f$ is the Coriolis parameter, $\rho_0$ is the sea water density, $\mathbf{\tau}_s$  and  $\mathbf{\tau}_b$ are the wind and bottom stress, respectively, and $\mathbf{X}$ is the internal mixing.

\section{Neural Networks fundamentals}
\label{sec:appendix_d}

\acrfull{ml} models exploit computers' capabilities of learning from past experiences (the input data) in order to make predictions. In this paper, we perform regression tasks, i.e., we predict the value of a continuous variable. These algorithms fall into the category of supervised learning because the models are trained with both features (input data) and labels (output data). For this, we use \acrfull{nn}, more specifically \acrfull{dnn}, a subclass of ML algorithms that use multiple layers to iteratively extract information from the training dataset. In each iteration, the signal travels from the input to the output layer, passing through the intermediate hidden layers. \acrshort{dnn} have the capability of modeling complex nonlinear relationships and are therefore suited for the problem we want to solve in this paper. A \acrshort{nn} has several components; all layers are conformed by nodes where the computation happens. The input data to each layer is combined with coefficients, also called weights, that either amplify or dampen the input. An activation function will then transform the weighted sum of the inputs to the neuron and pass on this information to the next layer or provide the predicted residual for the last layer. This way, different layers are able to apply different transformations to their inputs.

When we train a \acrshort{nn}, we adjust the weights to minimize the loss function, which traduces to reducing the MAE or the MSE:

\begin{equation}
    \mathrm{MAE} = \frac{1}{N}\sum_{i=1}^N | \mathrm{Predicted}_i - \mathrm{Observed}_i |, 
    \label{eq:mae}
\end{equation}

\begin{equation}
    \mathrm{MSE} = \frac{1}{N}\sum_{i=1}^N(\mathrm{Predicted}_i - \mathrm{Observed}_i)^2.
    \label{eq:mse}
\end{equation}

The training starts from random parameters updated for each learning iteration (epoch) to minimize the cost. The learning rate is one of the key parameters we have to tune; it defines how quickly we move toward a local minimum. A too large learning rate can overshoot, while a too small learning rate will require more iterations. Other important training parameters to consider in the design of the models' architecture are the number of layers and nodes per layer. However, it may only be feasible to test some combinations of parameters due to computational cost.

A popular metric for measuring the model performance is the \acrfull{rmse}:

\begin{equation}
    RMSE = \sqrt{MSE},
    \label{eq:rmse}
\end{equation}

The MSE can in turn be decompose into a bias and a variance component as follows:

\begin{equation}
    MSE = \mathrm{Bias}^2 + \mathrm{Var}.
\end{equation}

We aim to fit the input data by adjusting the learnable parameters in a model. Ideally, we want the models to capture the regularities in the training data and generalize well to unseen data. A model with high variance is characterized by high complexity and tends to overfit. In general, these models work well on the training data but fail to generalize on unseen data and therefore have a high test error. On the other hand, a model with high bias is too simple and unable to learn complex features. As a result, it underfits the data, fails to learn how to train the data, and has high training errors. The issue of overfitting can be addressed by: a) reducing the number of features (input data), b) using regularization, and c) early stopping. In this paper, we have used the three methods mentioned above. We have carefully selected the number of stations, variables, and range of hours used to train the model instead of using all possible variables. We have also used the dropout regularization method, which consists of randomly dropping out nodes during training. In addition, if no improvement is shown in the performance metric, the training will stop after 50 iterations.

\subsection{Number of predictors vs. number of samples}

In the field of \acrshort{ml}, the datasets are usually structured as tabular data, either in the form of arrays or dataframes. Most columns represent the predictor variables (also called input variables or features), while one column often represents the output (or labels). On the other hand, the rows are often referred to as samples (also called observations, records, or instances). Most \acrshort{ml} algorithms assume that the number of predictors ($p$) is much smaller than the number of samples ($n$): $p \ll n $. Therefore, as the dimension, $p$, increases, we need more samples to successfully represent the domain. 

Moreover, nonlinear models with higher flexibility and variance, like \acrshort{nn}s, depend more on the samples provided. For this reason, they require more data in the training process. Another factor that determines the amount of data needed is the complexity of the underlying function to learn. This is, we need enough data to capture the relationship between the predictors and between the predictors and the labels. 

Our study aims to predict the residuals in the numerical storm surge model, which is a nonlinear problem. We see that \acrshort{nn}s perform better than polar plots, but, as discussed, \acrshort{nn} are complex models that need more data than traditional statistical methods. Unfortunately, the number of samples in Nordic4-SS is limited, as the most recent version only has been run since 2018. The number of predictors is also limited, but it can quickly grow due to the number of stations, variables, and lagged hours. Despite the use of domain knowledge to reason about the data that may be needed to capture the complexity of the problem, feature selection is not a trivial task. In the following, we provide some numbers to illustrate the challenges associated with working with a small dataset.

Given that the labels are the residuals in Nordic4-SS, the number of predictions available from Nordic4-SS limits the number of samples in our problem. Nordic4-SS runs twice a day, meaning that, if no data is missing, from January 2018 to March 2021, the number of samples is $(365 \times 2 + 366 + 90) \times 2 = 2372$. Remember that we must split the data and leave some samples for testing. Due to seasonal variations, we have set apart an entire year for testing, resulting in a maximum of 1642 samples for training and 730 samples for test. When it comes to the predictors, we have explored the possibility of using observed total water level, tide estimates, storm surge predictions (also predictions generated in the past), pressure, wind, and wave data. If we limit the hour of past data to 24 hours before the analysis time and a lead time of 60 hours, we could potentially add data from 8 variables from the 23 permanent locations for 24 hours ($23 \times 8 \times 24 = 4416$ predictors), and for seven variables for +60 hours ($23 \times 7 \times 24 = 3864$ predictors). In addition, we can add forecasts from Nordic4-SS and MEPS generated 12 hours and 24 hours before the analysis time. As we see, the number of possible predictors is much greater than the number of samples and, because most of them a priori contain relevant information for the prediction of the residuals, it is not trivial which ones should be included in the \acrshort{ml} models. To limit the number of predictors, we reduced the number of stations from which we extract the parameters we will use to train the models, from 23 possible stations to 5, as explained in section \ref{sec:methods}. We also decided not to use past weather forecasts, and after running different experiments with winds, pressure, and waves, we discarded the pressure and wave data, as these do not improve results.


\subsection{Learning curves for the first lead time at Oscarsborg}

Learning curves of the model's performance are plots that show the progression over the experience of the learning in terms of a specific metric. They are widely used for algorithms that optimize their internal parameters incrementally over time, and we usually plot the training and the validation curves in the same figure to monitor whether the model is underfitting or overfitting. But the learning curves have multiple purposes in the learning process, such as diagnosing whether the train and validation datasets are representative of the problem domain, comparing different algorithms or choosing the model's parameters.

\begin{figure}
    \centering
    \includegraphics[scale=0.5]{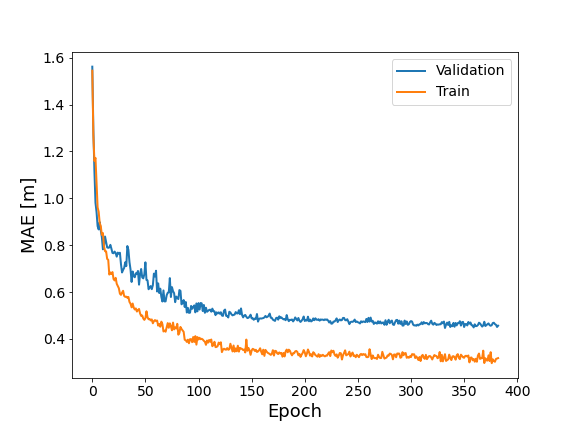}
    \caption{Mean Absolute Error (MAE)[m] as a function of epoch. The orange line shows the optimization learning curve for the training and the blue line the validation curve. The curves are obtained for a NN that learns the residuals in \acrshort{metno}'s storm surge model, Nordic4-SS, at the station \acrfull{osc} for lead time $t+1$. The training data corresponds to the period January 2018--March 2020 and test data to the period April 2020--March 2021.}
    \label{fig:learning_curves}
\end{figure}

The optimization learning curves for $t+1$ at \acrshort{osc} are illustrated in Fig. \ref{fig:learning_curves} and show how the algorithms learn incrementally from the data, as the error (MAE) is reduced with the number of iterations. The training curve provides an idea of how well the model is learning, while the validation curve indicates how well the model generalizes to previously unseen data. Therefore, we expect the training error to be lower than the validation error, as in Fig. \ref{fig:learning_curves}. Even though both the training and validation error decrease to a stability point, and the validation error is expected to be larger than the train error, a too wide gap between the lines could indicate unrepresentativeness due to a small training dataset. We will, however, show that the \acrshort{nn}s can improve Nordic4-SS. Moreover, when lead time increases (not shown), the gap between the curves increases and the convergence occurs for a smaller number of epochs. As expected, different stations have naturally different learning curves.

\section{Polar plots conditioned on wave parameters}
\label{sec:appendix_e}

\begin{figure}[ht]
  \centering
  \subfloat[Average residuals BGO]{\includegraphics[width=0.4\textwidth]{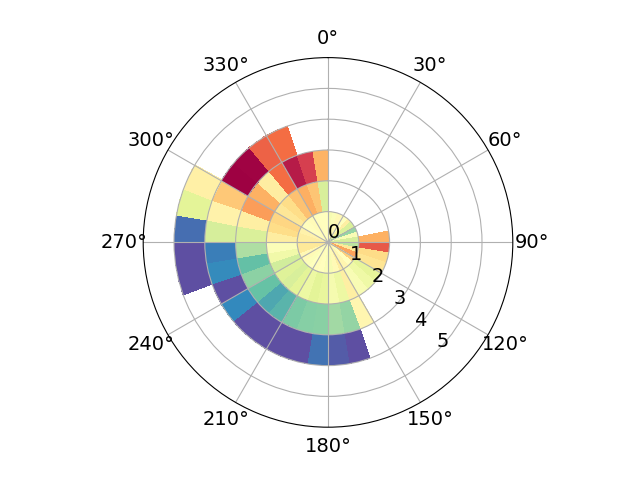}\label{fig:wave_radar_plot_mean_error_hindcast_NO_BGO}}
  \subfloat[Average residuals ANX]{\includegraphics[width=0.4\textwidth]{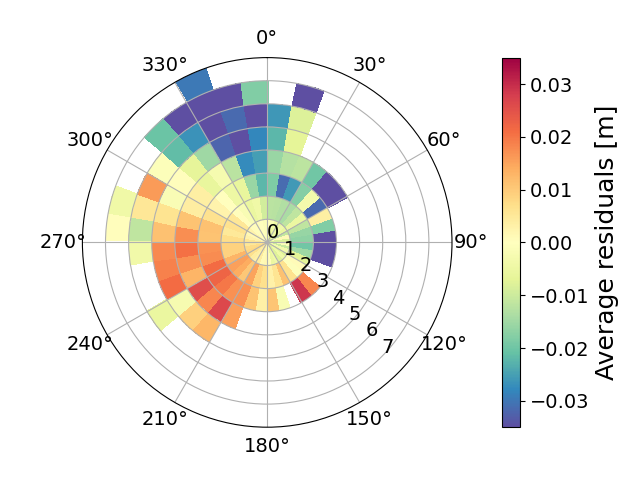}\label{fig:wave_radar_plot_mean_error_hindcast_NO_ANX}}
  
  \caption{Polar plots of average error in \acrshort{metno}'s storm surge hindcast (NORA-SS) significant wave height and mean wave direction. The residuals have been computed as the difference between the observed and predicted meteorological component, where the predicted meteorological component is the output from NORA-SS corrected with a weighted differences correction method. The panels show the average residuals at a) \acrfull{bgo}; and b) \acrfull{anx}. The bins are defined as boxes of size 1 m $\times$ 10 deg. Only bins with at least five observations are colored. Red colors indicate an underestimation and blue colors indicate an overestimation by the hindcast. The figures were constructed with hourly hindcast data from the period 2000--2019. Note that the
colorbars saturate for the highest values.}
  \label{fig:polar_plots_wave}
\end{figure}
Here, we show two examples of polar plots of the average residuals constructed for significant wave height in the radial coordinate and mean wave direction in the angular coordinate from ERA5 (Fig. \ref{fig:polar_plots_wave}). Fig. \ref{fig:wave_radar_plot_mean_error_hindcast_NO_BGO} shows that Nordic4-SS underestimates storm surges at \acrshort{bgo} for waves with mean direction $160^\circ - 280^\circ$, and overestimates for directions between $290^\circ - 360^\circ$. On the contrary, at \acrshort{anx}, Nordic4-SS overestimate surges with mean direction between $120^\circ - 300^\circ$, and underestimates for the other directions, as seen in Fig. \ref{fig:wave_radar_plot_mean_error_hindcast_NO_ANX}. We can also observe that significant wave heights are higher at \acrshort{anx} compared with \acrshort{bgo}. While the patterns in these figures appear clearly, we do not see any improvement by including both wind and wave predictors in the \acrshort{ml} model, presumably because waves are strongly correlated to winds as they are the result of the combination of wind and fetch. Since the wave data from ERA5 is not available at all locations, we decided to use the wind parameters as predictors.

\bibliographystyle{unsrt}
\bibliography{references}

\begin{thebibliography}{10}

\bibitem{pugh_1987}
David~T Pugh.
\newblock {\em Tides, surges and mean sea level}.
\newblock John Wiley and Sons Inc., New York, NY, 1987.

\bibitem{wood_2001}
Fergus~J Wood.
\newblock Tidal dynamics. {Volume II}: Extreme tidal peaks and coastal flooding.
\newblock {\em Journal of Coastal Research}, pages i--388, 2001.

\bibitem{gill_1982}
Adrian~E Gill.
\newblock {\em Atmosphere-ocean dynamics}, volume~30.
\newblock Academic Press, 1982.

\bibitem{ipccar6}
Hans-Otto P{\"o}rtner, Debra~C Roberts, H~Adams, C~Adler, P~Aldunce, E~Ali, R~Ara Begum, R~Betts, R~Bezner Kerr, R~Biesbroek, et~al.
\newblock Climate change 2022: Impacts, adaptation and vulnerability.
\newblock {\em IPCC Sixth Assessment Report}, 2022.

\bibitem{kristensen_et_al_2022}
Nils~Melsom Kristensen, Lars~Petter R{\o}ed, and {\O}yvind S{\ae}tra.
\newblock A forecasting and warning system of storm surge events along the norwegian coast.
\newblock {\em Environmental Fluid Mechanics}, pages 1--23, 2022.

\bibitem{lewis_et_al_2013}
M~Lewis, G~Schumann, P~Bates, and K~Horsburgh.
\newblock Understanding the variability of an extreme storm tide along a coastline.
\newblock {\em Estuarine, Coastal and Shelf Science}, 123:19--25, 2013.

\bibitem{mcinnes_et_al_2016}
Kathleen~L McInnes, Christopher~J White, Ivan~D Haigh, Mark~A Hemer, Ron~K Hoeke, Neil~J Holbrook, Anthony~S Kiem, Eric~CJ Oliver, Roshanka Ranasinghe, Kevin~JE Walsh, et~al.
\newblock {Natural hazards in Australia: sea level and coastal extremes}.
\newblock {\em Climatic Change}, 139(1):69--83, 2016.

\bibitem{hoffken_et_al_2020}
Jorid H{\"o}ffken, Athanasios~T Vafeidis, Leigh~R MacPherson, and S{\"o}nke Dangendorf.
\newblock Effects of the temporal variability of storm surges on coastal flooding.
\newblock {\em Frontiers in Marine Science}, 7:98, 2020.

\bibitem{tiggeloven_et_la_2021}
Timothy Tiggeloven, Ana{\"\i}s Couasnon, Chiem van Straaten, Sanne Muis, and Philip~J Ward.
\newblock Exploring deep learning capabilities for surge predictions in coastal areas.
\newblock {\em Scientific reports}, 11(1):1--15, 2021.

\bibitem{harris_1962}
D~Lee Harris.
\newblock The equivalence between certain statistical prediction methods and linearized dynamical methods.
\newblock {\em Monthly Weather Review}, 90(8):331--340, 1962.

\bibitem{pasquali_2020}
Davide Pasquali.
\newblock Simplified methods for storm surge forecast and hindcast in semi-enclosed basins: A review.
\newblock {\em Geophysics and Ocean Waves Studies}, 2020.

\bibitem{dramsch_2020}
Jesper~S{\"o}ren Dramsch.
\newblock 70 years of machine learning in geoscience in review.
\newblock {\em Advances in geophysics}, 61:1--55, 2020.

\bibitem{deoliveira_et_al_2009}
Marilia~MF De~Oliveira, Nelson Francisco~F Ebecken, Jorge Luiz~Fernandes De~Oliveira, and Isimar de~Azevedo~Santos.
\newblock Neural network model to predict a storm surge.
\newblock {\em Journal of applied Meteorology and Climatology}, 48(1):143--155, 2009.

\bibitem{kim_et_al_2016}
Sooyoul Kim, Yoshiharu Matsumi, Shunqi Pan, and Hajime Mase.
\newblock {A real-time forecast model using artificial neural network for after-runner storm surges on the Tottori coast, Japan}.
\newblock {\em Ocean Engineering}, 122:44--53, 2016.

\bibitem{kim_et_al_2019}
Sooyoul Kim, Shunqi Pan, and Hajime Mase.
\newblock {Artificial neural network-based storm surge forecast model: Practical application to Sakai Minato, Japan}.
\newblock {\em Applied Ocean Research}, 91:101871, 2019.

\bibitem{das_et_al_2011}
Himangshu~S Das, Hoonshin Jung, Bruce Ebersole, Ty~Wamsley, and Robert~W Whalin.
\newblock {An efficient storm surge forecasting tool for coastal Mississippi}.
\newblock {\em Coastal Engineering Proceedings}, 1(32):21--21, 2011.

\bibitem{tadesse_et_al_2020}
Michael Tadesse, Thomas Wahl, and A~Cid.
\newblock Data-driven modeling of global storm surges.
\newblock {\em Frontiers in Marine Science}, 7:260, 2020.

\bibitem{sztobryn_2003}
Marzenna Sztobryn.
\newblock Forecast of storm surge by means of artificial neural network.
\newblock {\em Journal of Sea Research}, 49(4):317--322, 2003.

\bibitem{makarynskyy_2004}
Oleg Makarynskyy, D~Makarynska, Michael Kuhn, and WE~Featherstone.
\newblock {Predicting sea level variations with artificial neural networks at Hillarys Boat Harbour, Western Australia}.
\newblock {\em Estuarine, Coastal and Shelf Science}, 61(2):351--360, 2004.

\bibitem{bruneau_et_al_2020}
Nicolas Bruneau, Jeff Polton, Joanne Williams, and Jason Holt.
\newblock Estimation of global coastal sea level extremes using neural networks.
\newblock {\em Environmental Research Letters}, 15(7):074030, 2020.

\bibitem{cox_et_al_2002}
Daniel~T Cox, Philippe Tissot, and Patrick Michaud.
\newblock Water level observations and short-term predictions including meteorological events for entrance of galveston bay, texas.
\newblock {\em Journal of waterway, port, coastal, and ocean engineering}, 128(1):21--29, 2002.

\bibitem{bajo_2010}
Marco Bajo and Georg Umgiesser.
\newblock Storm surge forecast through a combination of dynamic and neural network models.
\newblock {\em Ocean Modelling}, 33(1-2):1--9, 2010.

\bibitem{brantley_et_al_2013}
Halley Brantley, Gayle Hagler, Sue Kimbrough, R.~Williams, S.~Mukerjee, and Lucas Neas.
\newblock Mobile air monitoring data processing strategies and effects on spatial air pollution trends.
\newblock {\em Atmospheric Measurement Techniques}, 6:10443, 01 2013.

\bibitem{grange_et_al_2016}
Stuart~K. Grange, Alastair~C. Lewis, and David~C. Carslaw.
\newblock Source apportionment advances using polar plots of bivariate correlation and regression statistics.
\newblock {\em Atmospheric Environment}, 145:128--134, 2016.

\bibitem{carslaw_and_ropkins_2012}
David~C. Carslaw and Karl Ropkins.
\newblock openair --- an r package for air quality data analysis.
\newblock {\em Environmental Modelling \& Software}, 27--28(0):52--61, 2012.

\bibitem{api}
{The Norwegian Mapping Authority}.
\newblock {API for water level data}.
\newblock Web page, 2022.
\newblock Available online at: \url{https://api.sehavniva.no/tideapi_en.html}.

\bibitem{pytide}
{Centre National D'Études Spatiales}.
\newblock pangeo-pytide.
\newblock Web page, 2022.
\newblock Available online at: \url{https://github.com/CNES/pangeo-pytide}.

\bibitem{haidvogel_et_al_2008}
Dale~B Haidvogel, Hernan Arango, W~Paul Budgell, Bruce~D Cornuelle, Enrique Curchitser, Emanuele Di~Lorenzo, Katja Fennel, W~Rockwell Geyer, Albert~J Hermann, Lyon Lanerolle, et~al.
\newblock {Ocean forecasting in terrain-following coordinates: Formulation and skill assessment of the Regional Ocean Modeling System}.
\newblock {\em Journal of Computational Physics}, 227(7):3595--3624, 2008.

\bibitem{shchepetkin_and_mcwilliams_2005}
Alexander~F Shchepetkin and James~C McWilliams.
\newblock {The regional oceanic modeling system (ROMS): a split-explicit, free-surface, topography-following-coordinate oceanic model}.
\newblock {\em Ocean modelling}, 9(4):347--404, 2005.

\bibitem{api_met}
{Norwegian Meteorological Institute}.
\newblock {MET Norway Weather API}.
\newblock Web page, 2022.
\newblock Available online at: \url{https://api.met.no/}.

\bibitem{ocean_met}
{Norwegian Meteorological Institute}.
\newblock {ROMS}.
\newblock Web page, 2022.
\newblock Available online at: \url{https://ocean.met.no/models}.

\bibitem{engedahl_1995}
H~Engedahl.
\newblock Implementation of the princeton ocean model (pom/ecom-3d) at the norwegian meteorological institute (dnmi).
\newblock {\em Research Rep}, 5, 1995.

\bibitem{chapman_1985}
David~C Chapman.
\newblock Numerical treatment of cross-shelf open boundaries in a barotropic coastal ocean model.
\newblock {\em Journal of Physical oceanography}, 15(8):1060--1075, 1985.

\bibitem{flather_1976}
RA~Flather.
\newblock A tidal model of the northwest european continental shelf.
\newblock {\em Mem. Soc. Roy. Sci. Liege}, 10:141--164, 1976.

\bibitem{haidvogel_et_al_2000}
Dale~B Haidvogel, Hernan~G Arango, Kate Hedstrom, Aike Beckmann, Paola Malanotte-Rizzoli, and Alexander~F Shchepetkin.
\newblock {Model evaluation experiments in the North Atlantic Basin: simulations in nonlinear terrain-following coordinates}.
\newblock {\em Dynamics of atmospheres and oceans}, 32(3-4):239--281, 2000.

\bibitem{williams_et_al_2016}
Joanne Williams, Kevin~J Horsburgh, Jane~A Williams, and Robert~NF Proctor.
\newblock Tide and skew surge independence: New insights for flood risk.
\newblock {\em Geophysical Research Letters}, 43(12):6410--6417, 2016.

\bibitem{Bengtsson_et_al_2017}
Lisa Bengtsson, Ulf Andrae, Trygve Aspelien, Yurii Batrak, Javier Calvo, Wim de~Rooy, Emily Gleeson, Bent Hansen-Sass, Mariken Homleid, Mariano Hortal, et~al.
\newblock The harmonie--arome model configuration in the aladin--hirlam nwp system.
\newblock {\em Monthly Weather Review}, 145(5):1919--1935, 2017.

\bibitem{frogner_et_al_2019}
Inger-Lise Frogner, Ulf Andrae, Jelena Bojarova, Alfons Callado, PAU Escrib{\`a}, Henrik Feddersen, Alan Hally, Janne Kauhanen, Roger Randriamampianina, Andrew Singleton, et~al.
\newblock Harmoneps—the harmonie ensemble prediction system.
\newblock {\em Weather and Forecasting}, 34(6):1909--1937, 2019.

\bibitem{termonia_et_al_2018}
Piet Termonia, Claude Fischer, Eric Bazile, Fran{\c{c}}ois Bouyssel, Radmila Bro{\v{z}}kov{\'a}, Pierre B{\'e}nard, Bogdan Bochenek, Daan Degrauwe, Mari{\'a} Derkov{\'a}, Ryad El~Khatib, et~al.
\newblock The aladin system and its canonical model configurations arome cy41t1 and alaro cy40t1.
\newblock {\em Geoscientific Model Development}, 11(1):257--281, 2018.

\bibitem{toth_and_kalnay_1993}
Zoltan Toth and Eugenia Kalnay.
\newblock Ensemble forecasting at nmc: The generation of perturbations.
\newblock {\em Bulletin of the american meteorological society}, 74(12):2317--2330, 1993.

\bibitem{haakenstad_et_al_2021}
Hilde Haakenstad, \O{}yvind Breivik, Birgitte~R Furevik, Magnar Reistad, Patrik Bohlinger, and Ole~Johan Aarnes.
\newblock {NORA3: A nonhydrostatic high-resolution hindcast of the North Sea, the Norwegian Sea, and the Barents Sea}.
\newblock {\em Journal of Applied Meteorology and Climatology}, 2021.

\bibitem{solbrekke_et_al_2021}
I.~M. Solbrekke, A.~Sorteberg, and H.~Haakenstad.
\newblock {Norwegian hindcast archive (NORA3) -- A validation of offshore wind resources in the North Sea and Norwegian Sea}.
\newblock {\em Wind Energy Science Discussions}, 2021:1--31, 2021.

\bibitem{muller_et_al_2017_a}
Malte M{\"u}ller, Yurii Batrak, J{\o}rn Kristiansen, Morten~A{\O} K{\o}ltzow, Gunnar Noer, and Anton Korosov.
\newblock Characteristics of a convective-scale weather forecasting system for the european arctic.
\newblock {\em Monthly Weather Review}, 145(12):4771--4787, 2017.

\bibitem{muller_et_al_2017_b}
Malte M{\"u}ller, Mariken Homleid, Karl-Ivar Ivarsson, Morten~A{\O} K{\o}ltzow, Magnus Lindskog, Knut~Helge Midtb{\o}, Ulf Andrae, Trygve Aspelien, Lars Berggren, Dag Bj{\o}rge, et~al.
\newblock Arome-metcoop: A nordic convective-scale operational weather prediction model.
\newblock {\em Weather and Forecasting}, 32(2):609--627, 2017.

\bibitem{breivik_et_al_2022}
{\O}yvind Breivik, Ana Carrasco, Hilde Haakenstad, Ole~Johan Aarnes, Arno Behrens, Jean-Raymond Bidlot, Jan-Victor Bj{\"o}rkqvist, Patrik Bohlinger, Birgitte~R Furevik, Joanna Staneva, et~al.
\newblock The impact of a reduced high-wind charnock parameter on wave growth with application to the north sea, the norwegian sea, and the arctic ocean.
\newblock {\em Journal of Geophysical Research: Oceans}, 127(3):e2021JC018196, 2022.

\bibitem{hersbach_et_al_2019}
Hans Hersbach, W~Bell, P.~Berrisford, Andras Hor{\'a}nyi, Mu{\~n}oz-Sabater J., J.~Nicolas, Raluca Radu, Dinand Schepers, Adrian Simmons, Cornel Soci, and Dick Dee.
\newblock {Global reanalysis: goodbye ERA-Interim, hello ERA5}.
\newblock {\em ECMWF newsletter}, 159:17--24, 04 2019.

\bibitem{haakenstad_et_al_2022}
Hilde Haakenstad and {\O}yvind Breivik.
\newblock {NORA3. Part II: Precipitation and Temperature Statistics in Complex Terrain Modeled with a Nonhydrostatic Model}.
\newblock {\em Journal of Applied Meteorology and Climatology}, 61(10):1549--1572, 2022.

\bibitem{keras}
Fran\c{c}ois Chollet et~al.
\newblock Keras.
\newblock \url{https://github.com/fchollet/keras}, 2015.
\newblock Available online at: \url{https://github.com/fchollet/keras}.

\bibitem{pedregosa_et_al_2011}
F.~Pedregosa, G.~Varoquaux, A.~Gramfort, V.~Michel, B.~Thirion, O.~Grisel, M.~Blondel, P.~Prettenhofer, R.~Weiss, V.~Dubourg, J.~Vanderplas, A.~Passos, D.~Cournapeau, M.~Brucher, M.~Perrot, and E.~Duchesnay.
\newblock Scikit-learn: Machine learning in {P}ython.
\newblock {\em Journal of Machine Learning Research}, 12:2825--2830, 2011.

\bibitem{glorot_and_bengio_2010}
Xavier Glorot and Y.~Bengio.
\newblock Understanding the difficulty of training deep feedforward neural networks.
\newblock {\em Journal of Machine Learning Research - Proceedings Track}, 9:249--256, 01 2010.

\bibitem{kingma_and_ba_2014}
Diederik~P Kingma and Jimmy Ba.
\newblock Adam: A method for stochastic optimization.
\newblock {\em arXiv preprint arXiv:1412.6980}, 2014.

\bibitem{haigh_2017}
Ivan~D Haigh.
\newblock Tides and water levels.
\newblock {\em Encyclopedia of Maritime and Offshore Engineering}, pages 1--13, 2017.

\bibitem{kartverket}
{The Norwegian Mapping Authority}.
\newblock Se havniv\r{a}.
\newblock Web page, 2022.
\newblock Available online at: \url{https://www.kartverket.no/en/at-sea/se-havniva}.

\bibitem{resio_et_al_2008}
Donald~T. Resio and Joannes~J. Westerink.
\newblock {Modeling the physics of storm surges}.
\newblock {\em Physics Today}, 61(9):33--38, 09 2008.

\bibitem{kantha_and_clayson_2000}
Lakshmi~H Kantha and Carol~Anne Clayson.
\newblock {\em Numerical models of oceans and oceanic processes}.
\newblock Elsevier, 2000.

\bibitem{noaa}
{NOAA}.
\newblock {Storm Surge Overview}.
\newblock Web page, 2022.
\newblock Available online at: \url{https://www.nhc.noaa.gov/surge/}.

\end{thebibliography}

\end{document}